\documentclass[aps,prb,twocolumn,superscriptaddress,longbibliography]{revtex4-2}
\usepackage[pdftex, pdftitle={Near itinerancy and slow singlet formation in the triangular lattice NaRuO$_2$}, pdfauthor={Charles Tam}, colorlinks,allcolors=blue]{hyperref}
\usepackage{graphicx}
\usepackage{booktabs}
\usepackage{cleveref}
\usepackage{xcolor}

\begin{document}

\title{Near itinerancy and slow singlet formation in the triangular lattice NaRuO$_2$}

\author{Charles C. Tam}
\affiliation{Materials Department, University of California Santa Barbara, Santa Barbara, California 93106 USA}

\author{Alon Hendler Avidor}
\affiliation{Materials Department, University of California Santa Barbara, Santa Barbara, California 93106 USA}

\author{Pritam Bhattacharyya}
\affiliation{Institute for Theoretical Solid State Physics, Leibniz IFW Dresden, Helmholtzstraße 20, Dresden 01069, Germany}
\affiliation{SR University, Warangal, Telangana, 506371, India}

\author{Yongseong Choi}
\affiliation{Advanced Photon Source, Argonne National Laboratory, Argonne, Illinois 60439, USA}

\author{Daniel Haskel}
\affiliation{Advanced Photon Source, Argonne National Laboratory, Argonne, Illinois 60439, USA}

\author{Sven Luther}
\affiliation{Hochfeld-Magnetlabor Dresden (HLD-EMFL), Helmholtz-Zentrum Dresden-Rossendorf, 01328 Dresden, Germany}

\author{Hlynur Gretarsson}
\affiliation{Deutsches Elektronen-Synchrotron DESY, Hamburg D-22607, Germany}

\author{Liviu Hozoi}
\affiliation{Institute for Theoretical Solid State Physics, Leibniz IFW Dresden, Helmholtzstraße 20, Dresden 01069, Germany}

\author{Stephen D. Wilson}
\email{stephendwilson@ucsb.edu}
\affiliation{Materials Department, University of California Santa Barbara, Santa Barbara, California 93106 USA}

\begin{abstract}

NaRuO$_2$ forms a delafossite-like structure that contains triangular sublattices of edge-sharing RuO$_6$ octahedra. It shows no evidence of magnetic order down to 100~mK and persistent spin fluctuations, suggestive of a quantum disordered magnetic ground state. In order to characterize the physical regime from which this disordered state arises, we use resonant inelastic X-ray scattering (RIXS) and X-ray absorption spectroscopy (XAS) at the Ru-$L_{2,3}$-edge, along with pulsed high-field magnetization to characterize both the local electronic structure and the magnetic interactions. We infer significant spin-orbit coupling from XAS measurements. With RIXS, we observe intraorbital excitations, which are damped due to the near itinerancy of the band and suggestive of strong charge fluctuations. Corroborated by models of the high-field magnetization to a random singlet model, we propose a picture of a nearly itinerant system with strong magnetic and charge fluctuations that destabilize long-range magnetic order.


\end{abstract}

\maketitle

\section{Introduction}
Strong magnetic frustration on triangular lattice compounds has long been known to stabilize unconventional magnetic ground states, ranging from intrinsic quantum disordered states \cite{Anderson1973,Shimizu2003,Nakatsuji2005,Yamashita2008}, to spin supersolid states \cite{Momoi2000,Sengupta2007,Yamamoto2014}, to fascinating non-coplanar spin textures and spin vortex states~\cite{Okubo2012,Kurumaji2019,Park2023}.  The character of the electronic wave function decorating the triangular lattice is crucial to determining the ground state properties, ranging from isotropic spin only moments to spin-orbit entangled wave functions where strong anisotropies can be engineered. 

One of the more fascinating wavefunctions occurs in the form of $J_\mathrm{eff}=1/2$ spin-orbit entangled states built from $t_{2g}$ orbital manifolds. When they are placed in a network of edge-sharing octahedra, strong magnetic frustration results due to the relative amplification of anisotropic, bond-dependent exchange terms.  The seminal example of this occurs in spin-orbit assisted Mott insulators with $J_\mathrm{eff}=1/2$ pseudospins on a non-frustrated honeycomb lattice. The result is a dominant bond-dependent interaction known as Kitaev exchange, with realizations in materials such as $\alpha$-RuCl$_3$~\cite{Plumb2014,Sears2015}. In the limit where isotropic Heisenberg exchange vanishes, a spin liquid ground state is predicted~\cite{Kitaev2006,Jackeli2009,Chaloupka2010}, and there are proposals of unconventional Kitaev-driven states across a host of networks with edge-sharing octahedra \cite{Kimchi2014}, including triangular lattices~\cite{Li2015b}.


A competing instability that forms a quantum disordered ground state is the random singlet phase \cite{Shimokawa2015,Kimchi2018,Wu2019}.  In this scenario, residual disorder suppresses long-range magnetic order, such that the exchange is described by a probability distribution function, which may have a power-law behavior, e.g. $P(J) \sim J^{-\alpha}$. A number of spin liquid candidates have been shown to obey this behavior~\cite{Volkov2020,Khatua2022}, such as the case of YbMgGaO$_4$, where significant Mg-Ga disorder appears to suppress the magnetic order and is shown to form a random singlet state (RSS) rather than a quantum spin liquid (QSL)~\cite{Kimchi2018a}.

\begin{figure}[t]
\includegraphics[]{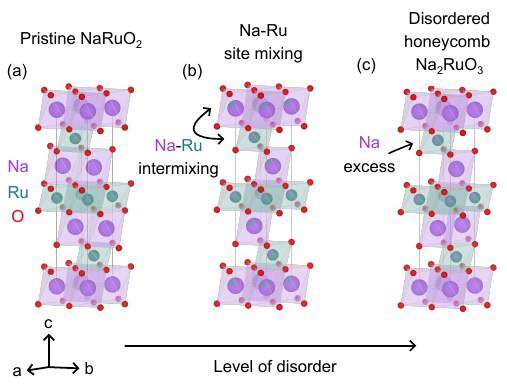}
\caption{\textbf{Possible defect modes in the triangular lattice of NaRuO$_2$.} (a) Drawing of the unit cell of pristine NaRuO$_2$. RuO$_6$ and NaO$_6$ octahedra have been drawn. (b) Illustration of a Na-Ru site mixing defect in NaRuO$_2$. (c) Illustration of disordered honeycomb Na$_2$RuO$_3$, which is equivalent to NaRuO$_2$ with a 1/3 occupation of Na on the Ru site.}
\label{fig:fig1} 
\end{figure}

An interesting triangular lattice magnet where both strong Kitaev exchange and the potential for disorder effects manifest is the compound NaRuO$_2$, where theoretical models report a large antiferromagnetic Kitaev term \cite{Razpopov2023} and Na/Ru site defects must be carefully managed \cite{Ortiz2022}.  NaRuO$_2$ is formed by stacked triangular lattices of NaO$_6$ and RuO$_6$ octahedra (\Cref{fig:fig1}a), and its magnetic ground state is disordered and characterized by slow, persistent fluctuations \cite{Ortiz2023}. The nature of its electronic ground state and the microscopic origin of these fluctuations remain open questions.

Specifically, in NaRuO$_2$, Ru naively assumes a $3^+$ oxidation state with a $d^5$ configuration, leading to a $J_\mathrm{eff}=1/2$ ground state~\cite{Razpopov2023}, analogous to $\alpha$-RuCl$_3$. While the electronic ground state is predicted to be insulating \cite{Razpopov2023, Bouhmouche2024}, consistent with electrical transport data \cite{Ortiz2023}, little is known about the character of the Mott phase. In particular, how strong the resulting Mott state is and the potential proximity of strong charge fluctuations remain undetermined experimentally. Equally undetermined is the impact of the dilute defects that persist in polycrystalline samples.  Illustrations of possible defect modes, as discussed in Ref.~\onlinecite{Ortiz2022}, are seen in \Cref{fig:fig1}b,c. Resolving the importance of charge fluctuations as well as the impact of dilute disorder on the magnetic ground state are important questions for resolving the nature of the quantum disordered ground state in NaRuO$_2$.


To clarify these open issues, here we report Ru-$L_{2,3}$-edge resonant inelastic X-ray scattering (RIXS), X-ray absorption spectroscopy (XAS) and pulsed high-field magnetization measurements on polycrystalline NaRuO$_2$ to characterize the nature of its electronic and magnetic ground states. While the SOC strength inferred from XAS is comparable to $\alpha$-RuCl$_3$, a clear $J_\mathrm{eff}=1/2$ excitation is not observed with RIXS, rather instead overdamped intraorbital excitations appear, reminiscent of a metal. This suggests the dominant role of charge fluctuations in NaRuO$_2$ relative to Kitaev terms in destabilizing magnetic order. We further show that the high-field, low-temperature magnetization can be captured with a random singlet model, where charge fluctuations associated with disorder from dilute Na-Ru intersite mixing~\cite{Ortiz2022} suppresses magnetic order.

\section{Methods}

Polycrystalline NaRuO$_2$ was synthesized by a solid-state reaction as detailed in Ref.~\onlinecite{Ortiz2023}. Polycrystalline samples were densified using spark plasma sintering and were polished to a mirror finish. Samples were measured without cleaving. Ru-$L_3$-edge RIXS was performed at the IRIXS spectrometer at P01, PETRA-III, DESY, with an energy resolution of around 70~meV~\cite{Gretarsson2020,Sundermann2026}. RIXS measurements were taken with a fixed incident angle of $\theta = 20^\circ$, a fixed scattering angle of $2\theta =90^\circ$ and at 20~K. Ru-$L_{2,3}$-edge XAS was performed at APS beamline 4ID, Argonne National Lab, and data were taken in fluorescence yield mode. 

XAS data were absorption corrected and normalized such that the $L_3$-edge absorption jump was unity. Spectra were fit to an arctangent broadened step function to fit the edge jump, and a Lorentzian peak. The center of the step and Lorentzian were constrained together, the width of the arctangent was fixed to tabulated values for the Ru-$L_{2,3}$-edges~\cite{Krause1979} while other parameters were left to float. We fit the RIXS spectra to a multipeak model, corresponding to the multiple inelastic features (phonons, and three higher energy modes A, B and C). The quasi-elastic peak is fit with a resolution-limited pseudo-Voigt function, and the phonon with a resolution-limited Gaussian. Mode A is fit to a damped harmonic (DHO) response, while mode B is well-fit with two Gaussians. The tail of mode C is fit with a linear background. The DHO response is given by

\begin{equation}\label{eq:dho}
    I(\omega) = A \frac{1}{1-\exp(-\;\hbar \omega/k_B T)} \frac{\omega_0}{(\omega^2-\omega_0^2) + \omega^2\gamma^2},
\end{equation}

where $\omega_0$ is the DHO pole frequency and $\gamma$ is the damping factor. With this definition, the DHO is damped when $\omega_0 > \gamma/2$ and overdamped when $\omega_0 < \gamma/2$.

Pulsed high-field magnetization measurements were conducted at the Dresden High Magnetic Field Laboratory, HLD-HZDR, using a magnetometer with a compensated pickup-coil system and a home-built $^3$He cryostat. Polycrystalline NaRuO$_2$ was packed in a quartz capillary in an Ar glovebox and flame-sealed. High-field measurements were calibrated to isothermal $M(H)$ curves taken with a Quantum design MPMS. Magnetic susceptibility and magnetization measurements were performed in a Quantum Design MPMS, on NaRuO$_2$ powder loaded in a brass holder.

Magnetic measurements were fit to a RSS model, which is detailed in Ref.~\onlinecite{Volkov2020}. In this model, the magnetization is given by 

\begin{equation}\label{eq:rs}
    M_\mathrm{RSS}(H) = \int_0^\infty \mathrm{d}J P(J) \frac{2 \mu \sinh(\beta \mu H)}{2 \cosh(\beta \mu H) + 1 + e^{\beta J}},
\end{equation}

where $\beta = 1/k_B T$, $H$ is the magnetic field, and $P(J) = \theta (J_0 - J)J^{-\alpha}$ is an empirical form of the exchange probability distribution function, where $\theta$ is a normalizing constant. Here, $\alpha$ is the power-law exponent and $J_0$ is a cutoff scale to avoid a diverging integral. We fix $J_0$ to 100~K for all fits. From Eq.~\ref{eq:rs}, it follows the magnetic susceptibility is given by $\chi_\mathrm{RS} = \partial M_\mathrm{RS}/\partial H$. We fit with a Van Vleck term, i.e. $\chi(T) = \chi_\mathrm{RS}(T) + \chi_0$, and $M(H) = M_\mathrm{RS}(H) + \chi_0 H$. Since fits are highly background dependent, the value of $\chi_0$ is constrained to be the same for all fits.

Multiplet-structure quantum chemical calculations were performed with focus on the branching ratio (BR), using the {\sc molpro} \cite{mo} package. Finite clusters and embeddings as described in refs.~\cite{Bhattacharyya2023,SR} were employed, to compute and compare BRs in 4$d^5$ NaRuO$_2$ and 4$d^5$ RuCl$_3$. The BRs were obtained following the procedure outlined in ref.~\cite{IC}, on the basis of complete-active-space self-consistent-field (CASSCF) computations \cite{qc,cas} for an average of the $^2T_{2g}$ $t_{2g}^5$, $^4T_{1g}$ and $^4T_{2g}$ $t_{2g}^4e_g^1$, $^6\!A_{1g}$ $t_{2g}^3e_g^2$, and low-lying $^2\!A_{2g}$, $^2T_{1g}$, $^2T_{2g}$, $^2\!E_g$ $t_{2g}^4e_g^1$ states.  In the subsequent spin-orbit treatment \cite{so}, all possible 4$d^5$ states were included.

\section{Results}

\subsection{Electronic structure}
\begin{figure}[t]
\includegraphics[]{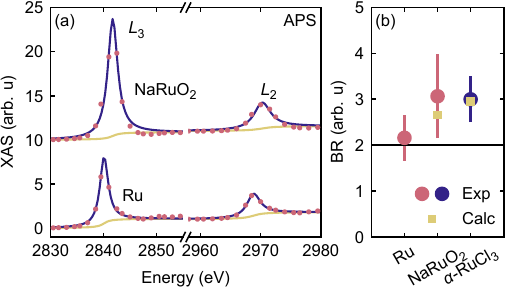}
\caption{\textbf{Branching ratios determined from Ru-$L$ XAS.} (a) XAS taken across the Ru $L_3$ and $L_2$ edges, on Ru and NaRuO$_2$. Data are fit to an arctangent step function and Lorentzian, defined in the methods. (b) Experimenmtal branching ratio (BR), defined as the fit intensity at $L_3$ divided by the intensity at $L_2$ for Ru, NaRuO$_2$, and $\alpha$-RuCl$_3$. The experimental BR for $\alpha$-RuCl$_3$ is taken from Ref.~\onlinecite{Plumb2014}. Details on the BR calculation for NaRuO$_2$ and $\alpha$-RuCl$_3$ are given in the methods. Energy resolution in the presented dataset was insufficient to identify finer detail in the $L_{3,2}$ peaks.}
\label{fig:fig3} 
\end{figure}

As an initial probe for the role of SOC in the ground state wave function of Ru$^{3+}$, XAS measurements were performed across the Ru-$L_{2,3}$-edges.  Here the Ru-$L_2$ edge corresponds to exciting $2p_{1/2} \rightarrow 4d$ while the $L_3$ edge is $2p_{3/2} \rightarrow 4d$. The angular part of the spin-orbit interaction $L \cdot S$ is related to the relative intensities of the $L_3$- and $L_2$-edge absorption peaks~\cite{Laan1988,Thole1988,Thole1988a,Laan2004} and is a conventional metric for classifying the strength of SOC in spin-orbit entangled wavefunctions~\cite{LagunaMarco2010,Clancy2012}. To investigate how the SOC in NaRuO$_2$ compares to other strongly spin-orbit entangled systems, XAS data for NaRuO$_2$ is plotted in \Cref{fig:fig3} and compared to Ru metal, as well as $\alpha$-RuCl$_3$~\cite{Plumb2014}. We note that generally, the BR for ruthenates is less than that of iridates. Taking the case of Sr$_2$IrO$_4$ and $\alpha$-RuCl$_3$, both have electron configuration $d^5$. The $5d$ states in the iridate are expected to have more SOC than in the ruthenate, since the SOC scales with $Z^2$ (account for screening). Accordingly, the measured SOC in Sr$_2$IrO$_4$ is around 0.4 eV~\cite{Zwartsenberg2020}, while in $\alpha$-RuCl$_3$ it is 0.15 eV~\cite{Suzuki2021}. Since the BR is impacted by the degree that SOC mixes $t_{2g}$ and $e_g$ states, it follows the BR for ruthenates with smaller SOC and larger U is less than iridates.\\

XAS data at the $L_3$ and $L_2$ edges were modeled as discussed in the methods section to determine their branching ratio (BR) (i.e. $I_{L_3}/I_{L_2}$). Fits are overplotted with the XAS spectra in \Cref{fig:fig3}a, and the extracted branching ratios for the two measured materials are plotted in \Cref{fig:fig3}b, alongside the BR for $\alpha$-RuCl$_3$ ~\cite{Plumb2014}. The BR for Ru metal is around 2, which is the expected statistical BR for a metal with a quenched spin-orbit interaction. In contrast, the BR value for NaRuO$_2$ is close to 3, similar to that found for the $J_\mathrm{eff}=1/2$ state in $\alpha$-RuCl$_3$ ~\cite{Plumb2014}. We also note similar BRs to $4d$ oxides with similar electron counts, namely $d^4$ SrRuO$_3$/SrTiO$_3$ films~\cite{Lamichhane2024}, and $d^5$ Li$_2$RhO$_3$~\cite{Bahrami2022}. This is consistent with the picture of a $J_\mathrm{eff}=1/2$ Mott state as the parent electronic phase for NaRuO$_2$. 

\begin{figure}[t]
\includegraphics[]{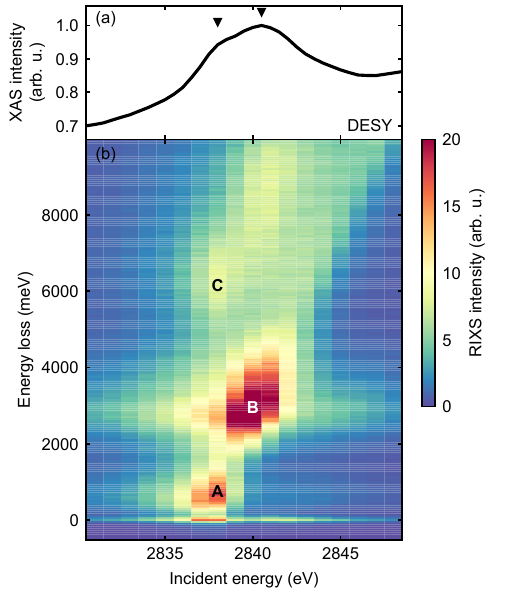}
\caption{\textbf{Ru-$L_3$-edge resonant inelastic X-ray scattering on polycrystalline NaRuO$_2$.} (a) XAS over the Ru $L_3$ edge, taken in fluorescence yield at room temperature. Marked peaks correspond to excitations into $t_{2g}$ and $e_g$ states. Normalization and self-absorption corrections have not been applied due to the small energy range collected. (b) RIXS energy map over the $L_3$ edge, showing three main features labelled A, B and C. RIXS data is momentum-averaged as the sample is polycrystalline, and taken at 20~K. }
\label{fig:fig2} 
\end{figure}

\begin{figure}[t]
\includegraphics[]{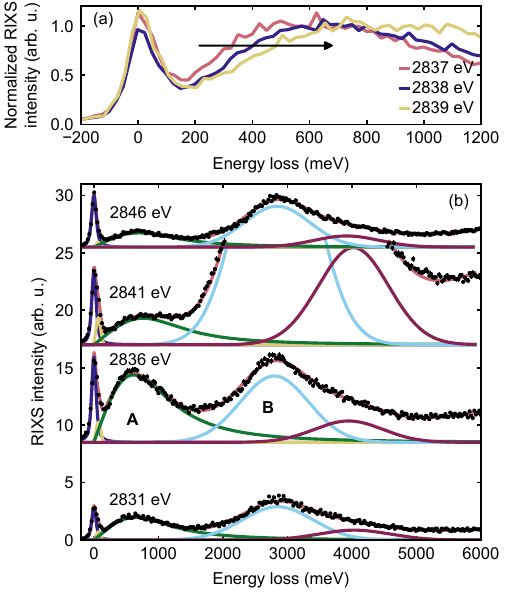}
\caption{\textbf{Energy dependent Ru-$L_3$-edge RIXS spectra.} (a) RIXS spectra normalized to intensity at 500~meV, across the $L_3$ edge. (b) RIXS spectra at various incident energies, showing their fits.}
\label{fig:fig22} 
\end{figure}

Quantum chemical computations yield a BR of 2.65 for NaRuO$_2$, somewhat lower than extracted from experiment, but still in the region corresponding to $J_{\mathrm{eff}}\!\approx\!1/2$ moments. Slight underestimation of the BR was also found in quantum chemical calculations on e.\,g. Ba$_2$IrO$_4$ \cite{IC}.
For comparison, a BR of 2.9 is obtained computationally for RuCl$_3$; experimentally, values of 3~\cite{Plumb2014} and 3.6 \cite{M_edge} were reported for the latter compound. The smaller BR for NaRuO$_2$ in the embedded-cluster quantum chemical computations can be understood on the basis of two effects\,: (i) a larger trigonal splitting ($\approx0.11$ eV in NaRuO$_2$ \cite{Bhattacharyya2023} vs $\approx0.07$ eV in RuCl$_3$ \cite{SR}), which yields a BR lower than the 2.75 value corresponding to pristine, cubic-environment $J_{\mathrm{eff}}\!=\!1/2$ moments with negligible configuration-interaction effects and 2nd-order SOCs within the $4d^5$ manifold;\,
(ii) a larger octahedral $t_{2g}$-$e_g$ splitting, with $t_{2g} \rightarrow e_g$ excitations starting
at $\approx1.6$ eV in NaRuO$_2$ (see \cite{Bhattacharyya2023} and next paragraphs) and at $\approx1.3$ eV in RuCl$_3$
\cite{SR,M_edge}, which makes configuration-interaction effects and 2nd-order SOCs somewhat less effective in NaRuO$_2$.


To explore the Mott state further, the low-energy Ru electronic states were studied by RIXS measurements at the Ru-$L_3$ edge. An overview of the RIXS data can be see in \Cref{fig:fig2}. To start, XAS data taken across the $L_3$ edge are plotted in \Cref{fig:fig2}a, where two peaks appear at 2838~eV and 2840.5~eV. Since NaRuO$_2$ possesses a $d^5$ filling, this corresponds to exciting into $t_{2g}$ and $e_g$ states. The energy separation of the two peaks is around 2.5~eV, which is consistent with the expected crystal field splitting. This is a similar value to similarly coordinated SrRuO$_3$~\cite{Zhang2022}.

\begin{figure}[t]
\includegraphics[]{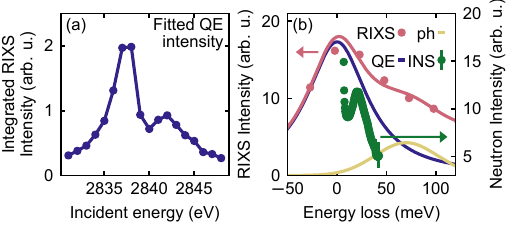}
\caption{\textbf{Low energy excitations in NaRuO$_2$.} (a) Fitted quasielastic RIXS intensity, as a function of incident energy. (b) RIXS response, taken at 2837~eV, where the datapoints are in red, fit is a red line, and quasielastic and phonon fit components are in blue and yellow. In green is the Q integrated energy cut inelastic neutron scattering data, taken at 1.8~K. The integration window is $[0,2]\;\mathrm{\AA}^{-1}$. Neutron data is from Ref.~\onlinecite{Ortiz2023}.}
\label{fig:fig5} 
\end{figure}

A RIXS intensity map as a function of incident photon energy and photon energy loss is plotted in Fig.~\ref{fig:fig2}b. Three main features  resonate across the $L_3$ edge, which we label features A, B, and C. In addition to these features, close to zero energy, a quasi-elastic peak and phonon mode appear. The RIXS spectra were fit to a model, defined in the methods section, that parameterizes the high energy electronic modes A, B, and C. Sample RIXS spectra collected at fixed incident energy, overplotted with their fit components, are presented in \Cref{fig:fig22}b.

Starting from high energy and working downward, feature C has the energy scale of a charge-transfer mode. Mode B is centered near 3~eV energy loss, suggestive of an interorbital excitation between $4d$ $t_{2g}$ and $e_g$ states. This corresponds well to interorbital excitations reported in other ruthenates 
(see e.g. Refs~\onlinecite{Takahashi2021,Gretarsson2024,Abdeldaim2024}). Finally, mode A appears near 500\;meV. Since this resonates at the $t_{2g}$ pre-peak, this mode potentially arises from intraorbital excitations within the $t_{2g}$ manifold, or, as discussed later, it represents a highly damped spin-orbit exciton. Modes A, B and C are all present across the Ru-$L_3$-edge, and all show moderate enhancement at resonance.

For most of the incident energy range, the fit DHO center and damping rate for mode A remain constant, with average values of $\omega_0 \approx 1000\;$meV and $\gamma \approx 1700\;$meV, meaning the mode is damped due to our definition of the DHO response in \Cref{eq:dho}. Near the $t_{2g}$ resonance, mode A starts moving to higher energy  and becomes overdamped at 2839~eV. This is more apparent when the spectra are normalized to mode A, as plotted in \Cref{fig:fig22}a. The mode is dispersing too slowly to be consistent with fluorescence, and instead the damped mode is reminiscent of a metal, where the local $t_{2g}$ transitions hybridize with the itinerant electrons to give a damped, slowly dispersing mode. 

Similar damped intraorbital excitations are observed in RIXS studies of other metallic ruthenates~\cite{Bertinshaw2021,Suzuki2023,Gretarsson2024}, and are not present in insulating ruthenates with a clearly defined gap~\cite{Gretarsson2019,Abdeldaim2024,Gretarsson2024}. There is no clearly defined excitation that can be assigned to a $J_\mathrm{eff}=1/2$ to $J_\mathrm{eff}=3/2$, as seen in insulating compounds with a $J_\mathrm{eff}=1/2$ ground state~\cite{kim2012,Suzuki2021}. However, we cannot rule out it is absent as it may be damped by the highly itinerant intraorbital fluctuations. This suggests that the Mott state in NaRuO$_2$ is weakened by the itinerancy of the system.

We also note the resonance effect of the quasielastic line. To see it more clearly, we plot the fitted intensity of the pseudovoigt used to model quasielastic scattering as a function of incident energy in \Cref{fig:fig5}a. We see a pronounced resonance effect peaked at $E_i \approx 2837\;$eV, which is the $t_{2g}$ resonance. Given the relatively poor energy resolution of $\Delta E \approx 70\;$meV, this would integrate many low energy features into the quasielastic scattering channel, e.g. phonons and spin fluctuations. Phonons in $4d$ systems tend to have a weaker resonance effect, and also tend to resonate at the higher energy $e_g$ resonance (c.f. Refs~\onlinecite{Gretarsson2019,Suzuki2021,Takahashi2021}). We can also preclude any static order in the charge channel, e.g. CDW, CDW fluctuations since the sample is polycrystalline. Therefore, it is likely these are low energy spin fluctuations that are being integrated into the quasielastic channel that are giving rise to the strong resonance effect. To compare with results from inelastic neutron scattering (INS) measurements, we plot a RIXS spectra at $E_i = 2837\;$eV along with a momentum integrated energy cut from Ref.~\onlinecite{Ortiz2023}, in \Cref{fig:fig5}b. A peak in the INS response at around 30~meV is consistent with being spin fluctuations~\cite{Ortiz2023}, which is in turn consistent with the energy scales of the quasielastic RIXS peak. This corroborates the presence of low energy spin fluctuations, which are present when measured with RIXS at 20~K, and are also present down to 1.8~K~\cite{Ortiz2023}.

\subsection{Magnetic Ground State}

\begin{figure*}[ht]
\includegraphics[]{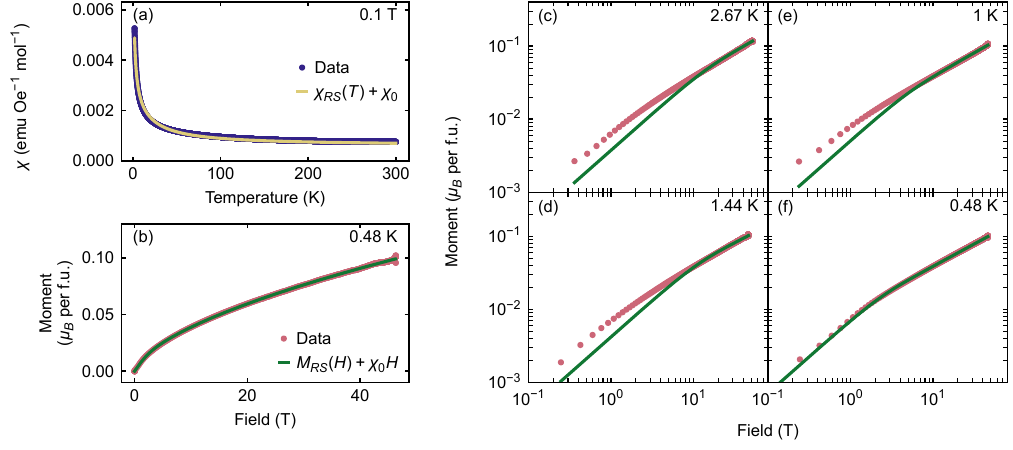}
\caption{\textbf{Random singlet fits to magnetic measurements of NaRuO$_2$.} Fits of the random singlet model to (a) magnetization taken at $T=0.48\;$K up to 45~T, and (b) magnetic susceptibility at $H=0.1\;$T. Definitions of $M_{RS}(H)$ and $\chi_{RS}(T)$ are given in the main text. (c)-(f) RSS fits to high-field magnetization at various temperatures on a log-log scale.}
\label{fig:fig4} 
\end{figure*}

Turning now to the origin of the low energy magnetic excitations in NaRuO$_2$, magnetization data were collected at low temperatures and under high magnetic fields.  The temperature dependence of magnetic susceptibility, taken under an applied field of $\mu_0 H=0.1\;$T is plotted in \Cref{fig:fig4}a. Consistent with earlier results, there is no evidence of magnetic ordering down to the lowest temperature measured of 1.8~K. Models of such a response, however, can be ambiguous.  Fits with a large Van Vleck-like constant susceptibility dressed with either a dilute concentration of paramagnetic impurities (parameterized as $\approx 2\%$ spin-$1/2$ moments from Ref.~\onlinecite{Ortiz2023}) or as a slow freezing of moments perform equally well. We note the inverse susceptibility does not fit a Curie-Weiss form, so obtaining an effective moment is not possible.

To gain deeper insight into the origin of the enhanced low-temperature susceptibility, isothermal high-field measurements were performed.  Pulsed high-field magnetization data taken at $T=0.48\;$K are plotted in \Cref{fig:fig4}b. Above $\approx$ 20~T, the slope of magnetization saturates, signifying a constant susceptibility. The total moment approaches around 0.1~$\mu_B$ per Ru ion at 45~T, consistent with the large magnetic exchange expected for this system. The lowest temperature 0.48~K isothermal data were fit to a random singlet RSS model \Cref{eq:rs}(\Cref{fig:fig4}f). This describes the data well, but quickly breaks down at 1~K and higher temperatures (\Cref{fig:fig4}c-e), which will be discussed more later. From the fits, the power-law exponent $\alpha$ can be extracted, which is summarized in \Cref{tab:tab1} and compared with the same RSS model fit to the temperature dependent susceptibility data in \Cref{fig:fig4}a.  The susceptibility of the RSS model was taken as the derivative of the magnetization \Cref{eq:rs}and the relatively large Van Vleck term of $\chi_0 = 5.9(2)\times10^{-4}$~emu~Oe~mol$^{-1}$ was kept fixed across all fits.  $\alpha$ values between fits yield good agreement between the $\chi(T)$ and $M(H)$ fits, with $\alpha \approx 0.4$ to 0.5. This suggests that the power-law scaling of the RSS exchange probability holds at the lowest temperature measured.

\begin{table}[ht]
\begin{tabular}{@{}cc@{}}
\toprule
Measurement & $\alpha$  \\ \midrule
$\chi(T)$ & 0.491(2) \\
$M(H,T=0.48\;\mathrm{K})$ & 0.434(1)  \\
$M(H,T=1\;\mathrm{K})$ & 0.393(2)   \\ \bottomrule
\end{tabular}
\caption{\textbf{RSS fit results.} Fit values for the power-law exponent $\alpha$ to the various datasets. The RSS model is defined in the methods.}
\label{tab:tab1} 
\end{table}

\section{Discussion and Conclusions}

Empirically, RIXS measurements suggest that NaRuO$_2$ is more itinerant than transport measurements suggest. As a result, the presence of a $J_\mathrm{eff}=1/2$ to $J_\mathrm{eff}=3/2$ exciton mode is not clear due to the strong damping from charge fluctuations. The SOC strength inferred from XAS is nevertheless comparable to $\alpha$-RuCl$_3$ and consistent with a $J_\mathrm{eff}=1/2$ state\; however we note that our XAS measurements lack the resolution to resolve the structure of the $L_2$ peak. We are therefore unable to comment of the nature of the $2p_{1/2} \rightarrow t_{2g}$ transition, which should be forbidden for a pure  $J_\mathrm{eff}=1/2$ ground state~\cite{Plumb2014}.

A good comparator of a Mott state in the metallic limit is Ru(Br$_{1-x}$I$_x$)$_3$. While RuBr$_3$ is a spin-orbit entangled Mott insulator~\cite{Imai2022}, RuI$_3$ is a semimetal~\cite{Nawa2021,Ni2022}, and a metal-insulator transition (MIT) occurs in intermediate compositions~\cite{Sato2024}. RIXS on these materials reveals a $J_\mathrm{eff}=1/2$ excitation that is present in both RuBr$_3$ and RuI$_3$, although it is more damped in RuI$_3$~\cite{Gretarsson2024}. This is noteable as this demonstrates the $J_\mathrm{eff}=1/2$ state can exist in a metal, although this case is distinct from NaRuO$_2$, as it is thought the bandwidth of the larger halogen $p$ orbitals drives the MIT~\cite{Gretarsson2024}, rather than the intrinsic weak Mott state in NaRuO$_2$ which we infer from our measurements.

Transport data reported earlier show a clear insulating/semiconducting behavior in $\approx 95$ $\%$ density samples \cite{Ortiz2023}, precluding trivial grain boundary effects from dominating the response.  Instead of a clean gap, variable-range hopping models better describe transport data, which are consistent with a polluted Mott gap and the low energy electronic excitations resolved in RIXS. Together, this suggests that the Mott state in NaRuO$_2$ is only marginally stable and that substantial charge fluctuations must be accounted for. 

Successful fits of magnetization data to an RSS model at the lowest measured temperature suggest that the Ru moments transition into a random singlet phase below $\approx1$ K.  This seemingly resolves an earlier puzzle of a small entropy anomaly and partial freezing previously observed near 1.5 K in heat capacity and susceptibility data \cite{Ortiz2023}.  While the small entropy change and weak cusp observed in susceptibility were consistent with a few percent fraction of free, impurity moments freezing, muon spin relaxation ($\mu$SR) measurements resolved the \textit{global} appearance of an inhomogeneous field distribution near this same temperature.  This suggests that the moments globally cross over into a new low-temperature phase below the 1.5 K energy scale, which is effectively probed by our high-field magnetization measurements at 0.5 K.  The agreement with the RSS model suggests that the onset of inhomogeneous spin fluctuations detected in $\mu$SR is the onset of slow singlet formation across the network of Ru moments.  This is potentially driven by the residual (several percent) chemical disorder likely present in the lattice, though we note that $\mu$SR also detects persistent, slow spin fluctuations in this low temperature state.

The synthesis of polycrystalline NaRuO$_2$ involves an excess of Ru which seems to mitigate the amount of Na-Ru site disorder. Fitting free spin impurity fractions from prior susceptibility and heat capacity data yields $\approx2\%$ molar fractions per formula unit, alongside a lower bound confidence for site mixing from neutron diffraction of around 5\%~\cite{Ortiz2023}. This fraction is substantially smaller than canonical RSS materials; however it can still dominate the magnetization signal relative to a quantum paramagnetic background.  Alternatively, in global models of a random singlet phase,  $J_2/J_1$--$\Delta$ models of disorder on an $S=1/2$ triangular lattice, predict a disordered spin phase, consistent with a RSS phase, as a fan above the QSL phase, for finite $\Delta > 0$~\cite{Wu2019}. In this picture of a global RSS phase, if NaRuO$_2$ is indeed proximate to a QSL in its pristine state, a small 2--5\% site mixing could be enough to destabilize the QSL into an RSS phase.

Above 1.5 K, fits to an RSS model break down, consistent with $\mu$SR's observation of dynamic moments in this regime \cite{Ortiz2023} and supportive of the notion of a crossover below this temperature.  Heat capacity data fit to different power-laws above and below this crossover, which preclude fits of C$_p$(T) to an RSS model. Higher field specific heat measurements ($\mu_0 H> 14$~T) are of interest for more fully shifting the crossover anomaly upward in temperature and obtaining a larger dynamic range for testing RSS fits.

Together our results paint a refined picture of the electronic state that gives rise to the unusual quantum disordered phase reported in NaRuO$_2$. Rather than a strongly local model involving strong Kitaev and anisotropic extended exchange terms with a fully gapped charge degree of freedom, RIXS and XAS data instead emphasize a starting point of a weak Mott phase with strong charge fluctuations.  While the starting $J_\mathrm{eff}=1/2$ wave functions are similar, NaRuO$_2$ is therefore distinct from $\alpha$-RuCl$_3$, likely due to its enhanced bandwidth. Instead, the strong role of charge fluctuations and the propensity of Ru to form dimers invites a closer comparison to organic Mott systems \cite{Kanoda2011}.  This weak Mott state paired with small levels of chemical disorder freezes into a state consistent with RSS models, though there is a clear crossover above 1.5 K into a dynamic spin state.  Whether or not this crossover is intrinsic or disorder-driven will require investigation of higher purity specimens, such as single-crystal samples. 


\section{Acknowledgments}

We acknowledge helpful discussions with Pavel Volkov. This work was supported by the US Department of Energy, Office of Science, Basic Energy Sciences under Award No. DE-SC0017752. Research was performed using facilities at the Advanced Photon Source, a U.S. Department of Energy (DOE) Office of Science user facility operated for the DOE Office of Science by Argonne National Laboratory under Contract No. DE-AC02-06CH11357.  We also acknowledge DESY – a member of the Helmholtz Association HGF – for access to beam time. We also acknowledge support of the HLD at HZDR, member of the European Magnetic Field Laboratory (EMFL).

\section{Data Availability}

The data that support the findings of this article are openly available \cite{zenodo}.


\begin{thebibliography}{62}%
\makeatletter
\providecommand \@ifxundefined [1]{%
 \@ifx{#1\undefined}
}%
\providecommand \@ifnum [1]{%
 \ifnum #1\expandafter \@firstoftwo
 \else \expandafter \@secondoftwo
 \fi
}%
\providecommand \@ifx [1]{%
 \ifx #1\expandafter \@firstoftwo
 \else \expandafter \@secondoftwo
 \fi
}%
\providecommand \natexlab [1]{#1}%
\providecommand \enquote  [1]{``#1''}%
\providecommand \bibnamefont  [1]{#1}%
\providecommand \bibfnamefont [1]{#1}%
\providecommand \citenamefont [1]{#1}%
\providecommand \href@noop [0]{\@secondoftwo}%
\providecommand \href [0]{\begingroup \@sanitize@url \@href}%
\providecommand \@href[1]{\@@startlink{#1}\@@href}%
\providecommand \@@href[1]{\endgroup#1\@@endlink}%
\providecommand \@sanitize@url [0]{\catcode `\\12\catcode `\$12\catcode `\&12\catcode `\#12\catcode `\^12\catcode `\_12\catcode `\%12\relax}%
\providecommand \@@startlink[1]{}%
\providecommand \@@endlink[0]{}%
\providecommand \url  [0]{\begingroup\@sanitize@url \@url }%
\providecommand \@url [1]{\endgroup\@href {#1}{\urlprefix }}%
\providecommand \urlprefix  [0]{URL }%
\providecommand \Eprint [0]{\href }%
\providecommand \doibase [0]{https://doi.org/}%
\providecommand \selectlanguage [0]{\@gobble}%
\providecommand \bibinfo  [0]{\@secondoftwo}%
\providecommand \bibfield  [0]{\@secondoftwo}%
\providecommand \translation [1]{[#1]}%
\providecommand \BibitemOpen [0]{}%
\providecommand \bibitemStop [0]{}%
\providecommand \bibitemNoStop [0]{.\EOS\space}%
\providecommand \EOS [0]{\spacefactor3000\relax}%
\providecommand \BibitemShut  [1]{\csname bibitem#1\endcsname}%
\let\auto@bib@innerbib\@empty
\bibitem [{\citenamefont {Anderson}(1973)}]{Anderson1973}%
  \BibitemOpen
  \bibfield  {author} {\bibinfo {author} {\bibfnamefont {P.}~\bibnamefont {Anderson}},\ }\bibfield  {title} {\bibinfo {title} {Resonating valence bonds: A new kind of insulator?},\ }\href {https://doi.org/10.1016/0025-5408(73)90167-0} {\bibfield  {journal} {\bibinfo  {journal} {Materials Research Bulletin}\ }\textbf {\bibinfo {volume} {8}},\ \bibinfo {pages} {153} (\bibinfo {year} {1973})}\BibitemShut {NoStop}%
\bibitem [{\citenamefont {Shimizu}\ \emph {et~al.}(2003)\citenamefont {Shimizu}, \citenamefont {Miyagawa}, \citenamefont {Kanoda}, \citenamefont {Maesato},\ and\ \citenamefont {Saito}}]{Shimizu2003}%
  \BibitemOpen
  \bibfield  {author} {\bibinfo {author} {\bibfnamefont {Y.}~\bibnamefont {Shimizu}}, \bibinfo {author} {\bibfnamefont {K.}~\bibnamefont {Miyagawa}}, \bibinfo {author} {\bibfnamefont {K.}~\bibnamefont {Kanoda}}, \bibinfo {author} {\bibfnamefont {M.}~\bibnamefont {Maesato}},\ and\ \bibinfo {author} {\bibfnamefont {G.}~\bibnamefont {Saito}},\ }\bibfield  {title} {\bibinfo {title} {Spin liquid state in an organic {Mott} insulator with a triangular lattice},\ }\href {https://doi.org/10.1103/physrevlett.91.107001} {\bibfield  {journal} {\bibinfo  {journal} {Physical Review Letters}\ }\textbf {\bibinfo {volume} {91}},\ \bibinfo {pages} {107001} (\bibinfo {year} {2003})}\BibitemShut {NoStop}%
\bibitem [{\citenamefont {Nakatsuji}\ \emph {et~al.}(2005)\citenamefont {Nakatsuji}, \citenamefont {Nambu}, \citenamefont {Tonomura}, \citenamefont {Sakai}, \citenamefont {Jonas}, \citenamefont {Broholm}, \citenamefont {Tsunetsugu}, \citenamefont {Qiu},\ and\ \citenamefont {Maeno}}]{Nakatsuji2005}%
  \BibitemOpen
  \bibfield  {author} {\bibinfo {author} {\bibfnamefont {S.}~\bibnamefont {Nakatsuji}}, \bibinfo {author} {\bibfnamefont {Y.}~\bibnamefont {Nambu}}, \bibinfo {author} {\bibfnamefont {H.}~\bibnamefont {Tonomura}}, \bibinfo {author} {\bibfnamefont {O.}~\bibnamefont {Sakai}}, \bibinfo {author} {\bibfnamefont {S.}~\bibnamefont {Jonas}}, \bibinfo {author} {\bibfnamefont {C.}~\bibnamefont {Broholm}}, \bibinfo {author} {\bibfnamefont {H.}~\bibnamefont {Tsunetsugu}}, \bibinfo {author} {\bibfnamefont {Y.}~\bibnamefont {Qiu}},\ and\ \bibinfo {author} {\bibfnamefont {Y.}~\bibnamefont {Maeno}},\ }\bibfield  {title} {\bibinfo {title} {Spin disorder on a triangular lattice},\ }\href {https://doi.org/10.1126/science.1114727} {\bibfield  {journal} {\bibinfo  {journal} {Science}\ }\textbf {\bibinfo {volume} {309}},\ \bibinfo {pages} {1697} (\bibinfo {year} {2005})}\BibitemShut {NoStop}%
\bibitem [{\citenamefont {Yamashita}\ \emph {et~al.}(2008)\citenamefont {Yamashita}, \citenamefont {Nakazawa}, \citenamefont {Oguni}, \citenamefont {Oshima}, \citenamefont {Nojiri}, \citenamefont {Shimizu}, \citenamefont {Miyagawa},\ and\ \citenamefont {Kanoda}}]{Yamashita2008}%
  \BibitemOpen
  \bibfield  {author} {\bibinfo {author} {\bibfnamefont {S.}~\bibnamefont {Yamashita}}, \bibinfo {author} {\bibfnamefont {Y.}~\bibnamefont {Nakazawa}}, \bibinfo {author} {\bibfnamefont {M.}~\bibnamefont {Oguni}}, \bibinfo {author} {\bibfnamefont {Y.}~\bibnamefont {Oshima}}, \bibinfo {author} {\bibfnamefont {H.}~\bibnamefont {Nojiri}}, \bibinfo {author} {\bibfnamefont {Y.}~\bibnamefont {Shimizu}}, \bibinfo {author} {\bibfnamefont {K.}~\bibnamefont {Miyagawa}},\ and\ \bibinfo {author} {\bibfnamefont {K.}~\bibnamefont {Kanoda}},\ }\bibfield  {title} {\bibinfo {title} {Thermodynamic properties of a spin-$1/2$ spin-liquid state in a $\kappa$-type organic salt},\ }\href {https://doi.org/10.1038/nphys942} {\bibfield  {journal} {\bibinfo  {journal} {Nature Physics}\ }\textbf {\bibinfo {volume} {4}},\ \bibinfo {pages} {459} (\bibinfo {year} {2008})}\BibitemShut {NoStop}%
\bibitem [{\citenamefont {Momoi}\ and\ \citenamefont {Totsuka}(2000)}]{Momoi2000}%
  \BibitemOpen
  \bibfield  {author} {\bibinfo {author} {\bibfnamefont {T.}~\bibnamefont {Momoi}}\ and\ \bibinfo {author} {\bibfnamefont {K.}~\bibnamefont {Totsuka}},\ }\bibfield  {title} {\bibinfo {title} {Magnetization plateaus of the {Shastry-Sutherland }model for {SrCu}$_2$({BO}$_3$)$_2$: Spin-density wave, supersolid, and bound states},\ }\href {https://doi.org/10.1103/physrevb.62.15067} {\bibfield  {journal} {\bibinfo  {journal} {Physical Review B}\ }\textbf {\bibinfo {volume} {62}},\ \bibinfo {pages} {15067} (\bibinfo {year} {2000})}\BibitemShut {NoStop}%
\bibitem [{\citenamefont {Sengupta}\ and\ \citenamefont {Batista}(2007)}]{Sengupta2007}%
  \BibitemOpen
  \bibfield  {author} {\bibinfo {author} {\bibfnamefont {P.}~\bibnamefont {Sengupta}}\ and\ \bibinfo {author} {\bibfnamefont {C.~D.}\ \bibnamefont {Batista}},\ }\bibfield  {title} {\bibinfo {title} {Field-induced supersolid phase in spin-one {Heisenberg} models},\ }\href {https://doi.org/10.1103/physrevlett.98.227201} {\bibfield  {journal} {\bibinfo  {journal} {Physical Review Letters}\ }\textbf {\bibinfo {volume} {98}},\ \bibinfo {pages} {227201} (\bibinfo {year} {2007})}\BibitemShut {NoStop}%
\bibitem [{\citenamefont {Yamamoto}\ \emph {et~al.}(2014)\citenamefont {Yamamoto}, \citenamefont {Marmorini},\ and\ \citenamefont {Danshita}}]{Yamamoto2014}%
  \BibitemOpen
  \bibfield  {author} {\bibinfo {author} {\bibfnamefont {D.}~\bibnamefont {Yamamoto}}, \bibinfo {author} {\bibfnamefont {G.}~\bibnamefont {Marmorini}},\ and\ \bibinfo {author} {\bibfnamefont {I.}~\bibnamefont {Danshita}},\ }\bibfield  {title} {\bibinfo {title} {Quantum phase diagram of the triangular-lattice {XXZ} model in a magnetic field},\ }\href {https://doi.org/10.1103/physrevlett.112.127203} {\bibfield  {journal} {\bibinfo  {journal} {Physical Review Letters}\ }\textbf {\bibinfo {volume} {112}},\ \bibinfo {pages} {127203} (\bibinfo {year} {2014})}\BibitemShut {NoStop}%
\bibitem [{\citenamefont {Okubo}\ \emph {et~al.}(2012)\citenamefont {Okubo}, \citenamefont {Chung},\ and\ \citenamefont {Kawamura}}]{Okubo2012}%
  \BibitemOpen
  \bibfield  {author} {\bibinfo {author} {\bibfnamefont {T.}~\bibnamefont {Okubo}}, \bibinfo {author} {\bibfnamefont {S.}~\bibnamefont {Chung}},\ and\ \bibinfo {author} {\bibfnamefont {H.}~\bibnamefont {Kawamura}},\ }\bibfield  {title} {\bibinfo {title} {Multiple-$q$ states and the {Skyrmion} lattice of the triangular-lattice heisenberg antiferromagnet under magnetic fields},\ }\href {https://doi.org/10.1103/physrevlett.108.017206} {\bibfield  {journal} {\bibinfo  {journal} {Physical Review Letters}\ }\textbf {\bibinfo {volume} {108}},\ \bibinfo {pages} {017206} (\bibinfo {year} {2012})}\BibitemShut {NoStop}%
\bibitem [{\citenamefont {Kurumaji}\ \emph {et~al.}(2019)\citenamefont {Kurumaji}, \citenamefont {Nakajima}, \citenamefont {Hirschberger}, \citenamefont {Kikkawa}, \citenamefont {Yamasaki}, \citenamefont {Sagayama}, \citenamefont {Nakao}, \citenamefont {Taguchi}, \citenamefont {Arima},\ and\ \citenamefont {Tokura}}]{Kurumaji2019}%
  \BibitemOpen
  \bibfield  {author} {\bibinfo {author} {\bibfnamefont {T.}~\bibnamefont {Kurumaji}}, \bibinfo {author} {\bibfnamefont {T.}~\bibnamefont {Nakajima}}, \bibinfo {author} {\bibfnamefont {M.}~\bibnamefont {Hirschberger}}, \bibinfo {author} {\bibfnamefont {A.}~\bibnamefont {Kikkawa}}, \bibinfo {author} {\bibfnamefont {Y.}~\bibnamefont {Yamasaki}}, \bibinfo {author} {\bibfnamefont {H.}~\bibnamefont {Sagayama}}, \bibinfo {author} {\bibfnamefont {H.}~\bibnamefont {Nakao}}, \bibinfo {author} {\bibfnamefont {Y.}~\bibnamefont {Taguchi}}, \bibinfo {author} {\bibfnamefont {T.-h.}\ \bibnamefont {Arima}},\ and\ \bibinfo {author} {\bibfnamefont {Y.}~\bibnamefont {Tokura}},\ }\bibfield  {title} {\bibinfo {title} {Skyrmion lattice with a giant topological {Hall} effect in a frustrated triangular-lattice magnet},\ }\href {https://doi.org/10.1126/science.aau0968} {\bibfield  {journal} {\bibinfo  {journal} {Science}\ }\textbf {\bibinfo {volume} {365}},\ \bibinfo {pages} {914} (\bibinfo {year} {2019})}\BibitemShut {NoStop}%
\bibitem [{\citenamefont {Park}\ \emph {et~al.}(2023)\citenamefont {Park}, \citenamefont {Cho}, \citenamefont {Kim}, \citenamefont {An}, \citenamefont {Kang}, \citenamefont {Avdeev}, \citenamefont {Sibille}, \citenamefont {Iida}, \citenamefont {Kajimoto}, \citenamefont {Lee}, \citenamefont {Ju}, \citenamefont {Cho}, \citenamefont {Noh}, \citenamefont {Han}, \citenamefont {Zhang}, \citenamefont {Batista},\ and\ \citenamefont {Park}}]{Park2023}%
  \BibitemOpen
  \bibfield  {author} {\bibinfo {author} {\bibfnamefont {P.}~\bibnamefont {Park}}, \bibinfo {author} {\bibfnamefont {W.}~\bibnamefont {Cho}}, \bibinfo {author} {\bibfnamefont {C.}~\bibnamefont {Kim}}, \bibinfo {author} {\bibfnamefont {Y.}~\bibnamefont {An}}, \bibinfo {author} {\bibfnamefont {Y.-G.}\ \bibnamefont {Kang}}, \bibinfo {author} {\bibfnamefont {M.}~\bibnamefont {Avdeev}}, \bibinfo {author} {\bibfnamefont {R.}~\bibnamefont {Sibille}}, \bibinfo {author} {\bibfnamefont {K.}~\bibnamefont {Iida}}, \bibinfo {author} {\bibfnamefont {R.}~\bibnamefont {Kajimoto}}, \bibinfo {author} {\bibfnamefont {K.~H.}\ \bibnamefont {Lee}}, \bibinfo {author} {\bibfnamefont {W.}~\bibnamefont {Ju}}, \bibinfo {author} {\bibfnamefont {E.-J.}\ \bibnamefont {Cho}}, \bibinfo {author} {\bibfnamefont {H.-J.}\ \bibnamefont {Noh}}, \bibinfo {author} {\bibfnamefont {M.~J.}\ \bibnamefont {Han}}, \bibinfo {author} {\bibfnamefont {S.-S.}\ \bibnamefont {Zhang}}, \bibinfo {author} {\bibfnamefont {C.~D.}\ \bibnamefont {Batista}},\ and\
  \bibinfo {author} {\bibfnamefont {J.-G.}\ \bibnamefont {Park}},\ }\bibfield  {title} {\bibinfo {title} {Tetrahedral triple-${Q}$ magnetic ordering and large spontaneous hall conductivity in the metallic triangular antiferromagnet {Co}$_{1/3}${TaS}$_2$},\ }\href {https://doi.org/10.1038/s41467-023-43853-4} {\bibfield  {journal} {\bibinfo  {journal} {Nature Communications}\ }\textbf {\bibinfo {volume} {14}},\ \bibinfo {pages} {8346} (\bibinfo {year} {2023})}\BibitemShut {NoStop}%
\bibitem [{\citenamefont {Plumb}\ \emph {et~al.}(2014)\citenamefont {Plumb}, \citenamefont {Clancy}, \citenamefont {Sandilands}, \citenamefont {Shankar}, \citenamefont {Hu}, \citenamefont {Burch}, \citenamefont {Kee},\ and\ \citenamefont {Kim}}]{Plumb2014}%
  \BibitemOpen
  \bibfield  {author} {\bibinfo {author} {\bibfnamefont {K.~W.}\ \bibnamefont {Plumb}}, \bibinfo {author} {\bibfnamefont {J.~P.}\ \bibnamefont {Clancy}}, \bibinfo {author} {\bibfnamefont {L.~J.}\ \bibnamefont {Sandilands}}, \bibinfo {author} {\bibfnamefont {V.~V.}\ \bibnamefont {Shankar}}, \bibinfo {author} {\bibfnamefont {Y.~F.}\ \bibnamefont {Hu}}, \bibinfo {author} {\bibfnamefont {K.~S.}\ \bibnamefont {Burch}}, \bibinfo {author} {\bibfnamefont {H.-Y.}\ \bibnamefont {Kee}},\ and\ \bibinfo {author} {\bibfnamefont {Y.-J.}\ \bibnamefont {Kim}},\ }\bibfield  {title} {\bibinfo {title} {$\alpha$-{RuCl}$_3$: A spin-orbit assisted {Mott} insulator on a honeycomb lattice},\ }\href {https://doi.org/10.1103/physrevb.90.041112} {\bibfield  {journal} {\bibinfo  {journal} {Physical Review B}\ }\textbf {\bibinfo {volume} {90}},\ \bibinfo {pages} {041112} (\bibinfo {year} {2014})}\BibitemShut {NoStop}%
\bibitem [{\citenamefont {Sears}\ \emph {et~al.}(2015)\citenamefont {Sears}, \citenamefont {Songvilay}, \citenamefont {Plumb}, \citenamefont {Clancy}, \citenamefont {Qiu}, \citenamefont {Zhao}, \citenamefont {Parshall},\ and\ \citenamefont {Kim}}]{Sears2015}%
  \BibitemOpen
  \bibfield  {author} {\bibinfo {author} {\bibfnamefont {J.~A.}\ \bibnamefont {Sears}}, \bibinfo {author} {\bibfnamefont {M.}~\bibnamefont {Songvilay}}, \bibinfo {author} {\bibfnamefont {K.~W.}\ \bibnamefont {Plumb}}, \bibinfo {author} {\bibfnamefont {J.~P.}\ \bibnamefont {Clancy}}, \bibinfo {author} {\bibfnamefont {Y.}~\bibnamefont {Qiu}}, \bibinfo {author} {\bibfnamefont {Y.}~\bibnamefont {Zhao}}, \bibinfo {author} {\bibfnamefont {D.}~\bibnamefont {Parshall}},\ and\ \bibinfo {author} {\bibfnamefont {Y.-J.}\ \bibnamefont {Kim}},\ }\bibfield  {title} {\bibinfo {title} {Magnetic order in $\alpha$-{RuCl}$_3$: A honeycomb-lattice quantum magnet with strong spin-orbit coupling},\ }\href {https://doi.org/10.1103/physrevb.91.144420} {\bibfield  {journal} {\bibinfo  {journal} {Physical Review B}\ }\textbf {\bibinfo {volume} {91}},\ \bibinfo {pages} {144420} (\bibinfo {year} {2015})}\BibitemShut {NoStop}%
\bibitem [{\citenamefont {Kitaev}(2006)}]{Kitaev2006}%
  \BibitemOpen
  \bibfield  {author} {\bibinfo {author} {\bibfnamefont {A.}~\bibnamefont {Kitaev}},\ }\bibfield  {title} {\bibinfo {title} {Anyons in an exactly solved model and beyond},\ }\href {https://doi.org/10.1016/j.aop.2005.10.005} {\bibfield  {journal} {\bibinfo  {journal} {Annals of Physics}\ }\textbf {\bibinfo {volume} {321}},\ \bibinfo {pages} {2} (\bibinfo {year} {2006})}\BibitemShut {NoStop}%
\bibitem [{\citenamefont {Jackeli}\ and\ \citenamefont {Khaliullin}(2009)}]{Jackeli2009}%
  \BibitemOpen
  \bibfield  {author} {\bibinfo {author} {\bibfnamefont {G.}~\bibnamefont {Jackeli}}\ and\ \bibinfo {author} {\bibfnamefont {G.}~\bibnamefont {Khaliullin}},\ }\bibfield  {title} {\bibinfo {title} {Mott insulators in the strong spin-orbit coupling limit: From {Heisenberg} to a quantum compass and {Kitaev} models},\ }\href {https://doi.org/10.1103/physrevlett.102.017205} {\bibfield  {journal} {\bibinfo  {journal} {Physical Review Letters}\ }\textbf {\bibinfo {volume} {102}},\ \bibinfo {pages} {017205} (\bibinfo {year} {2009})}\BibitemShut {NoStop}%
\bibitem [{\citenamefont {Chaloupka}\ \emph {et~al.}(2010)\citenamefont {Chaloupka}, \citenamefont {Jackeli},\ and\ \citenamefont {Khaliullin}}]{Chaloupka2010}%
  \BibitemOpen
  \bibfield  {author} {\bibinfo {author} {\bibfnamefont {J.}~\bibnamefont {Chaloupka}}, \bibinfo {author} {\bibfnamefont {G.}~\bibnamefont {Jackeli}},\ and\ \bibinfo {author} {\bibfnamefont {G.}~\bibnamefont {Khaliullin}},\ }\bibfield  {title} {\bibinfo {title} {Kitaev-heisenberg model on a honeycomb lattice: Possible exotic phases in iridium oxides {A}$_2${IrO}$_3$},\ }\href {https://doi.org/10.1103/physrevlett.105.027204} {\bibfield  {journal} {\bibinfo  {journal} {Physical Review Letters}\ }\textbf {\bibinfo {volume} {105}},\ \bibinfo {pages} {027204} (\bibinfo {year} {2010})}\BibitemShut {NoStop}%
\bibitem [{\citenamefont {Kimchi}\ and\ \citenamefont {Vishwanath}(2014)}]{Kimchi2014}%
  \BibitemOpen
  \bibfield  {author} {\bibinfo {author} {\bibfnamefont {I.}~\bibnamefont {Kimchi}}\ and\ \bibinfo {author} {\bibfnamefont {A.}~\bibnamefont {Vishwanath}},\ }\bibfield  {title} {\bibinfo {title} {{Kitaev-Heisenberg} models for iridates on the triangular, hyperkagome, kagome, fcc, and pyrochlore lattices},\ }\href {https://doi.org/10.1103/physrevb.89.014414} {\bibfield  {journal} {\bibinfo  {journal} {Physical Review B}\ }\textbf {\bibinfo {volume} {89}},\ \bibinfo {pages} {014414} (\bibinfo {year} {2014})}\BibitemShut {NoStop}%
\bibitem [{\citenamefont {Li}\ \emph {et~al.}(2015)\citenamefont {Li}, \citenamefont {Yu},\ and\ \citenamefont {Li}}]{Li2015b}%
  \BibitemOpen
  \bibfield  {author} {\bibinfo {author} {\bibfnamefont {K.}~\bibnamefont {Li}}, \bibinfo {author} {\bibfnamefont {S.-L.}\ \bibnamefont {Yu}},\ and\ \bibinfo {author} {\bibfnamefont {J.-X.}\ \bibnamefont {Li}},\ }\bibfield  {title} {\bibinfo {title} {Global phase diagram, possible chiral spin liquid, and topological superconductivity in the triangular {Kitaev-Heisenberg} model},\ }\href {https://doi.org/10.1088/1367-2630/17/4/043032} {\bibfield  {journal} {\bibinfo  {journal} {New Journal of Physics}\ }\textbf {\bibinfo {volume} {17}},\ \bibinfo {pages} {043032} (\bibinfo {year} {2015})}\BibitemShut {NoStop}%
\bibitem [{\citenamefont {Shimokawa}\ \emph {et~al.}(2015)\citenamefont {Shimokawa}, \citenamefont {Watanabe},\ and\ \citenamefont {Kawamura}}]{Shimokawa2015}%
  \BibitemOpen
  \bibfield  {author} {\bibinfo {author} {\bibfnamefont {T.}~\bibnamefont {Shimokawa}}, \bibinfo {author} {\bibfnamefont {K.}~\bibnamefont {Watanabe}},\ and\ \bibinfo {author} {\bibfnamefont {H.}~\bibnamefont {Kawamura}},\ }\bibfield  {title} {\bibinfo {title} {Static and dynamical spin correlations of the $s=1/2$ random-bond antiferromagnetic {Heisenberg} model on the triangular and kagome lattices},\ }\href {https://doi.org/10.1103/physrevb.92.134407} {\bibfield  {journal} {\bibinfo  {journal} {Physical Review B}\ }\textbf {\bibinfo {volume} {92}},\ \bibinfo {pages} {134407} (\bibinfo {year} {2015})}\BibitemShut {NoStop}%
\bibitem [{\citenamefont {Kimchi}\ \emph {et~al.}(2018{\natexlab{a}})\citenamefont {Kimchi}, \citenamefont {Sheckelton}, \citenamefont {McQueen},\ and\ \citenamefont {Lee}}]{Kimchi2018}%
  \BibitemOpen
  \bibfield  {author} {\bibinfo {author} {\bibfnamefont {I.}~\bibnamefont {Kimchi}}, \bibinfo {author} {\bibfnamefont {J.~P.}\ \bibnamefont {Sheckelton}}, \bibinfo {author} {\bibfnamefont {T.~M.}\ \bibnamefont {McQueen}},\ and\ \bibinfo {author} {\bibfnamefont {P.~A.}\ \bibnamefont {Lee}},\ }\bibfield  {title} {\bibinfo {title} {Scaling and data collapse from local moments in frustrated disordered quantum spin systems},\ }\href {https://doi.org/10.1038/s41467-018-06800-2} {\bibfield  {journal} {\bibinfo  {journal} {Nature Communications}\ }\textbf {\bibinfo {volume} {9}},\ \bibinfo {pages} {4367} (\bibinfo {year} {2018}{\natexlab{a}})}\BibitemShut {NoStop}%
\bibitem [{\citenamefont {Wu}\ \emph {et~al.}(2019)\citenamefont {Wu}, \citenamefont {Gong},\ and\ \citenamefont {Sheng}}]{Wu2019}%
  \BibitemOpen
  \bibfield  {author} {\bibinfo {author} {\bibfnamefont {H.-Q.}\ \bibnamefont {Wu}}, \bibinfo {author} {\bibfnamefont {S.-S.}\ \bibnamefont {Gong}},\ and\ \bibinfo {author} {\bibfnamefont {D.~N.}\ \bibnamefont {Sheng}},\ }\bibfield  {title} {\bibinfo {title} {Randomness-induced spin-liquid-like phase in the spin-$1/2$ $j_1$-$j_2$ triangular {Heisenberg} model},\ }\href {https://doi.org/10.1103/physrevb.99.085141} {\bibfield  {journal} {\bibinfo  {journal} {Physical Review B}\ }\textbf {\bibinfo {volume} {99}},\ \bibinfo {pages} {085141} (\bibinfo {year} {2019})}\BibitemShut {NoStop}%
\bibitem [{\citenamefont {Volkov}\ \emph {et~al.}(2020)\citenamefont {Volkov}, \citenamefont {Won}, \citenamefont {Gorbunov}, \citenamefont {Kim}, \citenamefont {Ye}, \citenamefont {Kim}, \citenamefont {Pixley}, \citenamefont {Cheong},\ and\ \citenamefont {Blumberg}}]{Volkov2020}%
  \BibitemOpen
  \bibfield  {author} {\bibinfo {author} {\bibfnamefont {P.~A.}\ \bibnamefont {Volkov}}, \bibinfo {author} {\bibfnamefont {C.-J.}\ \bibnamefont {Won}}, \bibinfo {author} {\bibfnamefont {D.~I.}\ \bibnamefont {Gorbunov}}, \bibinfo {author} {\bibfnamefont {J.}~\bibnamefont {Kim}}, \bibinfo {author} {\bibfnamefont {M.}~\bibnamefont {Ye}}, \bibinfo {author} {\bibfnamefont {H.-S.}\ \bibnamefont {Kim}}, \bibinfo {author} {\bibfnamefont {J.~H.}\ \bibnamefont {Pixley}}, \bibinfo {author} {\bibfnamefont {S.-W.}\ \bibnamefont {Cheong}},\ and\ \bibinfo {author} {\bibfnamefont {G.}~\bibnamefont {Blumberg}},\ }\bibfield  {title} {\bibinfo {title} {Random singlet state in {Ba}$_5${CuIr}$_3${O}$_{12}$ single crystals},\ }\href {https://doi.org/10.1103/physrevb.101.020406} {\bibfield  {journal} {\bibinfo  {journal} {Physical Review B}\ }\textbf {\bibinfo {volume} {101}},\ \bibinfo {pages} {020406} (\bibinfo {year} {2020})}\BibitemShut {NoStop}%
\bibitem [{\citenamefont {Khatua}\ \emph {et~al.}(2022)\citenamefont {Khatua}, \citenamefont {Gomilsek}, \citenamefont {Orain}, \citenamefont {Strydom}, \citenamefont {Jaglicic}, \citenamefont {Colin}, \citenamefont {Petit}, \citenamefont {Ozarowski}, \citenamefont {Mangin-Thro}, \citenamefont {Sethupathi}, \citenamefont {Rao}, \citenamefont {Zorko},\ and\ \citenamefont {Khuntia}}]{Khatua2022}%
  \BibitemOpen
  \bibfield  {author} {\bibinfo {author} {\bibfnamefont {J.}~\bibnamefont {Khatua}}, \bibinfo {author} {\bibfnamefont {M.}~\bibnamefont {Gomilsek}}, \bibinfo {author} {\bibfnamefont {J.~C.}\ \bibnamefont {Orain}}, \bibinfo {author} {\bibfnamefont {A.~M.}\ \bibnamefont {Strydom}}, \bibinfo {author} {\bibfnamefont {Z.}~\bibnamefont {Jaglicic}}, \bibinfo {author} {\bibfnamefont {C.~V.}\ \bibnamefont {Colin}}, \bibinfo {author} {\bibfnamefont {S.}~\bibnamefont {Petit}}, \bibinfo {author} {\bibfnamefont {A.}~\bibnamefont {Ozarowski}}, \bibinfo {author} {\bibfnamefont {L.}~\bibnamefont {Mangin-Thro}}, \bibinfo {author} {\bibfnamefont {K.}~\bibnamefont {Sethupathi}}, \bibinfo {author} {\bibfnamefont {M.~S.~R.}\ \bibnamefont {Rao}}, \bibinfo {author} {\bibfnamefont {A.}~\bibnamefont {Zorko}},\ and\ \bibinfo {author} {\bibfnamefont {P.}~\bibnamefont {Khuntia}},\ }\bibfield  {title} {\bibinfo {title} {Signature of a randomness-driven spin-liquid state in a frustrated magnet},\ }\href
  {https://doi.org/10.1038/s42005-022-00879-2} {\bibfield  {journal} {\bibinfo  {journal} {Communications Physics}\ }\textbf {\bibinfo {volume} {5}},\ \bibinfo {pages} {99} (\bibinfo {year} {2022})}\BibitemShut {NoStop}%
\bibitem [{\citenamefont {Kimchi}\ \emph {et~al.}(2018{\natexlab{b}})\citenamefont {Kimchi}, \citenamefont {Nahum},\ and\ \citenamefont {Senthil}}]{Kimchi2018a}%
  \BibitemOpen
  \bibfield  {author} {\bibinfo {author} {\bibfnamefont {I.}~\bibnamefont {Kimchi}}, \bibinfo {author} {\bibfnamefont {A.}~\bibnamefont {Nahum}},\ and\ \bibinfo {author} {\bibfnamefont {T.}~\bibnamefont {Senthil}},\ }\bibfield  {title} {\bibinfo {title} {Valence bonds in random quantum magnets: Theory and application to {YbMgGaO}$_4$},\ }\href {https://doi.org/10.1103/physrevx.8.031028} {\bibfield  {journal} {\bibinfo  {journal} {Physical Review X}\ }\textbf {\bibinfo {volume} {8}},\ \bibinfo {pages} {031028} (\bibinfo {year} {2018}{\natexlab{b}})}\BibitemShut {NoStop}%
\bibitem [{\citenamefont {Razpopov}\ \emph {et~al.}(2023)\citenamefont {Razpopov}, \citenamefont {Kaib}, \citenamefont {Backes}, \citenamefont {Balents}, \citenamefont {Wilson}, \citenamefont {Ferrari}, \citenamefont {Riedl},\ and\ \citenamefont {Valenti}}]{Razpopov2023}%
  \BibitemOpen
  \bibfield  {author} {\bibinfo {author} {\bibfnamefont {A.}~\bibnamefont {Razpopov}}, \bibinfo {author} {\bibfnamefont {D.~A.~S.}\ \bibnamefont {Kaib}}, \bibinfo {author} {\bibfnamefont {S.}~\bibnamefont {Backes}}, \bibinfo {author} {\bibfnamefont {L.}~\bibnamefont {Balents}}, \bibinfo {author} {\bibfnamefont {S.~D.}\ \bibnamefont {Wilson}}, \bibinfo {author} {\bibfnamefont {F.}~\bibnamefont {Ferrari}}, \bibinfo {author} {\bibfnamefont {K.}~\bibnamefont {Riedl}},\ and\ \bibinfo {author} {\bibfnamefont {R.}~\bibnamefont {Valenti}},\ }\bibfield  {title} {\bibinfo {title} {A ${J}_{eff}=1/2$ {Kitaev }material on the trianglur lattice: the case of {NaRuO}$_2$},\ }\href {https://doi.org/10.1038/s41535-023-00567-6} {\bibfield  {journal} {\bibinfo  {journal} {npj Quantum Materials}\ }\textbf {\bibinfo {volume} {8}},\ \bibinfo {pages} {36} (\bibinfo {year} {2023})}\BibitemShut {NoStop}%
\bibitem [{\citenamefont {Ortiz}\ \emph {et~al.}(2022)\citenamefont {Ortiz}, \citenamefont {Sarte}, \citenamefont {Avidor},\ and\ \citenamefont {Wilson}}]{Ortiz2022}%
  \BibitemOpen
  \bibfield  {author} {\bibinfo {author} {\bibfnamefont {B.~R.}\ \bibnamefont {Ortiz}}, \bibinfo {author} {\bibfnamefont {P.~M.}\ \bibnamefont {Sarte}}, \bibinfo {author} {\bibfnamefont {A.~H.}\ \bibnamefont {Avidor}},\ and\ \bibinfo {author} {\bibfnamefont {S.~D.}\ \bibnamefont {Wilson}},\ }\bibfield  {title} {\bibinfo {title} {Defect control in the {Heisenberg-Kitaev} candidate material {NaRuO}$_2$},\ }\href {https://doi.org/10.1103/physrevmaterials.6.104413} {\bibfield  {journal} {\bibinfo  {journal} {Physical Review Materials}\ }\textbf {\bibinfo {volume} {6}},\ \bibinfo {pages} {104413} (\bibinfo {year} {2022})}\BibitemShut {NoStop}%
\bibitem [{\citenamefont {Ortiz}\ \emph {et~al.}(2023)\citenamefont {Ortiz}, \citenamefont {Sarte}, \citenamefont {Avidor}, \citenamefont {Hay}, \citenamefont {Kenney}, \citenamefont {Kolesnikov}, \citenamefont {Pajerowski}, \citenamefont {Aczel}, \citenamefont {Taddei}, \citenamefont {Brown}, \citenamefont {Wang}, \citenamefont {Graf}, \citenamefont {Seshadri}, \citenamefont {Balents},\ and\ \citenamefont {Wilson}}]{Ortiz2023}%
  \BibitemOpen
  \bibfield  {author} {\bibinfo {author} {\bibfnamefont {B.~R.}\ \bibnamefont {Ortiz}}, \bibinfo {author} {\bibfnamefont {P.~M.}\ \bibnamefont {Sarte}}, \bibinfo {author} {\bibfnamefont {A.~H.}\ \bibnamefont {Avidor}}, \bibinfo {author} {\bibfnamefont {A.}~\bibnamefont {Hay}}, \bibinfo {author} {\bibfnamefont {E.}~\bibnamefont {Kenney}}, \bibinfo {author} {\bibfnamefont {A.~I.}\ \bibnamefont {Kolesnikov}}, \bibinfo {author} {\bibfnamefont {D.~M.}\ \bibnamefont {Pajerowski}}, \bibinfo {author} {\bibfnamefont {A.~A.}\ \bibnamefont {Aczel}}, \bibinfo {author} {\bibfnamefont {K.~M.}\ \bibnamefont {Taddei}}, \bibinfo {author} {\bibfnamefont {C.~M.}\ \bibnamefont {Brown}}, \bibinfo {author} {\bibfnamefont {C.}~\bibnamefont {Wang}}, \bibinfo {author} {\bibfnamefont {M.~J.}\ \bibnamefont {Graf}}, \bibinfo {author} {\bibfnamefont {R.}~\bibnamefont {Seshadri}}, \bibinfo {author} {\bibfnamefont {L.}~\bibnamefont {Balents}},\ and\ \bibinfo {author} {\bibfnamefont {S.~D.}\ \bibnamefont {Wilson}},\ }\bibfield  {title}
  {\bibinfo {title} {Quantum disordered ground state in the triangular-lattice magnet {NaRuO}$_2$},\ }\href {https://doi.org/10.1038/s41567-023-02039-x} {\bibfield  {journal} {\bibinfo  {journal} {Nature Physics}\ }\textbf {\bibinfo {volume} {19}},\ \bibinfo {pages} {943} (\bibinfo {year} {2023})}\BibitemShut {NoStop}%
\bibitem [{\citenamefont {Bouhmouche}\ \emph {et~al.}(2024)\citenamefont {Bouhmouche}, \citenamefont {Rhrissi}, \citenamefont {Jabar},\ and\ \citenamefont {Moubah}}]{Bouhmouche2024}%
  \BibitemOpen
  \bibfield  {author} {\bibinfo {author} {\bibfnamefont {A.}~\bibnamefont {Bouhmouche}}, \bibinfo {author} {\bibfnamefont {I.}~\bibnamefont {Rhrissi}}, \bibinfo {author} {\bibfnamefont {A.}~\bibnamefont {Jabar}},\ and\ \bibinfo {author} {\bibfnamefont {R.}~\bibnamefont {Moubah}},\ }\bibfield  {title} {\bibinfo {title} {Electronic, optical, and thermodynamic properties of a quantum spin liquid candidate {NaRuO}$_2$ ab-initio investigation},\ }\href {https://doi.org/10.1016/j.matchemphys.2023.128860} {\bibfield  {journal} {\bibinfo  {journal} {Materials Chemistry and Physics}\ }\textbf {\bibinfo {volume} {314}},\ \bibinfo {pages} {128860} (\bibinfo {year} {2024})}\BibitemShut {NoStop}%
\bibitem [{\citenamefont {Gretarsson}\ \emph {et~al.}(2020)\citenamefont {Gretarsson}, \citenamefont {Ketenoglu}, \citenamefont {Harder}, \citenamefont {Mayer}, \citenamefont {Dill}, \citenamefont {Spiwek}, \citenamefont {Schulte-Schrepping}, \citenamefont {Tischer}, \citenamefont {Wille}, \citenamefont {Keimer},\ and\ \citenamefont {Yavaş}}]{Gretarsson2020}%
  \BibitemOpen
  \bibfield  {author} {\bibinfo {author} {\bibfnamefont {H.}~\bibnamefont {Gretarsson}}, \bibinfo {author} {\bibfnamefont {D.}~\bibnamefont {Ketenoglu}}, \bibinfo {author} {\bibfnamefont {M.}~\bibnamefont {Harder}}, \bibinfo {author} {\bibfnamefont {S.}~\bibnamefont {Mayer}}, \bibinfo {author} {\bibfnamefont {F.-U.}\ \bibnamefont {Dill}}, \bibinfo {author} {\bibfnamefont {M.}~\bibnamefont {Spiwek}}, \bibinfo {author} {\bibfnamefont {H.}~\bibnamefont {Schulte-Schrepping}}, \bibinfo {author} {\bibfnamefont {M.}~\bibnamefont {Tischer}}, \bibinfo {author} {\bibfnamefont {H.-C.}\ \bibnamefont {Wille}}, \bibinfo {author} {\bibfnamefont {B.}~\bibnamefont {Keimer}},\ and\ \bibinfo {author} {\bibfnamefont {H.}~\bibnamefont {Yavaş}},\ }\bibfield  {title} {\bibinfo {title} {{IRIXS}: a resonant inelastic x-ray scattering instrument dedicated to x-rays in the intermediate energy range},\ }\href {https://doi.org/10.1107/s1600577519017119} {\bibfield  {journal} {\bibinfo  {journal} {Journal of Synchrotron Radiation}\
  }\textbf {\bibinfo {volume} {27}},\ \bibinfo {pages} {538} (\bibinfo {year} {2020})}\BibitemShut {NoStop}%
\bibitem [{\citenamefont {Sundermann}\ \emph {et~al.}(2026)\citenamefont {Sundermann}, \citenamefont {Harder}, \citenamefont {Said}, \citenamefont {Keimer},\ and\ \citenamefont {Gretarsson}}]{Sundermann2026}%
  \BibitemOpen
  \bibfield  {author} {\bibinfo {author} {\bibfnamefont {M.}~\bibnamefont {Sundermann}}, \bibinfo {author} {\bibfnamefont {M.}~\bibnamefont {Harder}}, \bibinfo {author} {\bibfnamefont {A.~H.}\ \bibnamefont {Said}}, \bibinfo {author} {\bibfnamefont {B.}~\bibnamefont {Keimer}},\ and\ \bibinfo {author} {\bibfnamefont {H.}~\bibnamefont {Gretarsson}},\ }\bibfield  {title} {\bibinfo {title} {Dispersion-compensated rowland spectrometer: implications for uranium {VB-RIXS}},\ }\href {https://doi.org/10.1107/s1600577525010318} {\bibfield  {journal} {\bibinfo  {journal} {Journal of Synchrotron Radiation}\ }\textbf {\bibinfo {volume} {33}},\ \bibinfo {pages} {218} (\bibinfo {year} {2026})}\BibitemShut {NoStop}%
\bibitem [{\citenamefont {Krause}\ and\ \citenamefont {Oliver}(1979)}]{Krause1979}%
  \BibitemOpen
  \bibfield  {author} {\bibinfo {author} {\bibfnamefont {M.~O.}\ \bibnamefont {Krause}}\ and\ \bibinfo {author} {\bibfnamefont {J.~H.}\ \bibnamefont {Oliver}},\ }\bibfield  {title} {\bibinfo {title} {Natural widths of atomic {K} and {L} levels, {K}$\alpha$ {X}-ray lines and several {KLL} {A}uger lines},\ }\href {https://doi.org/10.1063/1.555595} {\bibfield  {journal} {\bibinfo  {journal} {Journal of Physical and Chemical Reference Data}\ }\textbf {\bibinfo {volume} {8}},\ \bibinfo {pages} {329} (\bibinfo {year} {1979})}\BibitemShut {NoStop}%
\bibitem [{\citenamefont {Werner}\ \emph {et~al.}(2012)\citenamefont {Werner}, \citenamefont {Knowles}, \citenamefont {Knizia}, \citenamefont {Manby},\ and\ \citenamefont {Sch{\"u}tz}}]{mo}%
  \BibitemOpen
  \bibfield  {author} {\bibinfo {author} {\bibfnamefont {H.-J.}\ \bibnamefont {Werner}}, \bibinfo {author} {\bibfnamefont {P.~J.}\ \bibnamefont {Knowles}}, \bibinfo {author} {\bibfnamefont {G.}~\bibnamefont {Knizia}}, \bibinfo {author} {\bibfnamefont {F.~R.}\ \bibnamefont {Manby}},\ and\ \bibinfo {author} {\bibfnamefont {M.}~\bibnamefont {Sch{\"u}tz}},\ }\bibfield  {title} {\bibinfo {title} {Molpro: a general-purpose quantum chemistry program package},\ }\href {https://doi.org/https://doi.org/10.1002/wcms.82} {\bibfield  {journal} {\bibinfo  {journal} {WIREs Comput. Mol. Sci.}\ }\textbf {\bibinfo {volume} {2}},\ \bibinfo {pages} {242} (\bibinfo {year} {2012})}\BibitemShut {NoStop}%
\bibitem [{\citenamefont {Bhattacharyya}\ \emph {et~al.}(2023)\citenamefont {Bhattacharyya}, \citenamefont {Bogdanov}, \citenamefont {Nishimoto}, \citenamefont {Wilson},\ and\ \citenamefont {Hozoi}}]{Bhattacharyya2023}%
  \BibitemOpen
  \bibfield  {author} {\bibinfo {author} {\bibfnamefont {P.}~\bibnamefont {Bhattacharyya}}, \bibinfo {author} {\bibfnamefont {N.~A.}\ \bibnamefont {Bogdanov}}, \bibinfo {author} {\bibfnamefont {S.}~\bibnamefont {Nishimoto}}, \bibinfo {author} {\bibfnamefont {S.~D.}\ \bibnamefont {Wilson}},\ and\ \bibinfo {author} {\bibfnamefont {L.}~\bibnamefont {Hozoi}},\ }\bibfield  {title} {\bibinfo {title} {{NaRuO}$_2$: {Kitaev-Heisenberg} exchange in triangular-lattice setting},\ }\href {https://doi.org/10.1038/s41535-023-00582-7} {\bibfield  {journal} {\bibinfo  {journal} {npj Quantum Materials}\ }\textbf {\bibinfo {volume} {8}},\ \bibinfo {pages} {52} (\bibinfo {year} {2023})}\BibitemShut {NoStop}%
\bibitem [{\citenamefont {Yadav}\ \emph {et~al.}(2016)\citenamefont {Yadav}, \citenamefont {Bogdanov}, \citenamefont {Katukuri}, \citenamefont {Nishimoto}, \citenamefont {van~den Brink},\ and\ \citenamefont {Hozoi}}]{SR}%
  \BibitemOpen
  \bibfield  {author} {\bibinfo {author} {\bibfnamefont {R.}~\bibnamefont {Yadav}}, \bibinfo {author} {\bibfnamefont {N.~A.}\ \bibnamefont {Bogdanov}}, \bibinfo {author} {\bibfnamefont {V.~M.}\ \bibnamefont {Katukuri}}, \bibinfo {author} {\bibfnamefont {S.}~\bibnamefont {Nishimoto}}, \bibinfo {author} {\bibfnamefont {J.}~\bibnamefont {van~den Brink}},\ and\ \bibinfo {author} {\bibfnamefont {L.}~\bibnamefont {Hozoi}},\ }\bibfield  {title} {\bibinfo {title} {Kitaev exchange and field-induced quantum spin-liquid states in honeycomb $\alpha$-{RuCl}$_3$},\ }\href {https://doi.org/10.1038/srep37925} {\bibfield  {journal} {\bibinfo  {journal} {Sci. Rep.}\ }\textbf {\bibinfo {volume} {6}},\ \bibinfo {pages} {37925} (\bibinfo {year} {2016})}\BibitemShut {NoStop}%
\bibitem [{\citenamefont {Katukuri}\ \emph {et~al.}(2014)\citenamefont {Katukuri}, \citenamefont {Roszeitis}, \citenamefont {Yushankhai}, \citenamefont {Mitrushchenkov}, \citenamefont {Stoll}, \citenamefont {van Veenendaal}, \citenamefont {Fulde}, \citenamefont {van~den Brink},\ and\ \citenamefont {Hozoi}}]{IC}%
  \BibitemOpen
  \bibfield  {author} {\bibinfo {author} {\bibfnamefont {V.~M.}\ \bibnamefont {Katukuri}}, \bibinfo {author} {\bibfnamefont {K.}~\bibnamefont {Roszeitis}}, \bibinfo {author} {\bibfnamefont {V.}~\bibnamefont {Yushankhai}}, \bibinfo {author} {\bibfnamefont {A.}~\bibnamefont {Mitrushchenkov}}, \bibinfo {author} {\bibfnamefont {H.}~\bibnamefont {Stoll}}, \bibinfo {author} {\bibfnamefont {M.}~\bibnamefont {van Veenendaal}}, \bibinfo {author} {\bibfnamefont {P.}~\bibnamefont {Fulde}}, \bibinfo {author} {\bibfnamefont {J.}~\bibnamefont {van~den Brink}},\ and\ \bibinfo {author} {\bibfnamefont {L.}~\bibnamefont {Hozoi}},\ }\bibfield  {title} {\bibinfo {title} {{Electronic Structure of Low-Dimensional 4d$^5$ Oxides: Interplay of Ligand Distortions, Overall Lattice Anisotropy, and Spin-Orbit Interactions}},\ }\href {https://doi.org/10.1021/ic402653f} {\bibfield  {journal} {\bibinfo  {journal} {Inorg. Chem.}\ }\textbf {\bibinfo {volume} {53}},\ \bibinfo {pages} {4833} (\bibinfo {year} {2014})}\BibitemShut {NoStop}%
\bibitem [{\citenamefont {Helgaker}\ \emph {et~al.}(2000)\citenamefont {Helgaker}, \citenamefont {J{\o}rgensen},\ and\ \citenamefont {Olsen}}]{qc}%
  \BibitemOpen
  \bibfield  {author} {\bibinfo {author} {\bibfnamefont {T.}~\bibnamefont {Helgaker}}, \bibinfo {author} {\bibfnamefont {P.}~\bibnamefont {J{\o}rgensen}},\ and\ \bibinfo {author} {\bibfnamefont {J.}~\bibnamefont {Olsen}},\ }\href@noop {} {\emph {\bibinfo {title} {Molecular Electronic Structure Theory}}}\ (\bibinfo  {publisher} {John Wiley \& Sons},\ \bibinfo {address} {Chichester},\ \bibinfo {year} {2000})\BibitemShut {NoStop}%
\bibitem [{\citenamefont {Kreplin}\ \emph {et~al.}(2020)\citenamefont {Kreplin}, \citenamefont {Knowles},\ and\ \citenamefont {Werner}}]{cas}%
  \BibitemOpen
  \bibfield  {author} {\bibinfo {author} {\bibfnamefont {D.~A.}\ \bibnamefont {Kreplin}}, \bibinfo {author} {\bibfnamefont {P.~J.}\ \bibnamefont {Knowles}},\ and\ \bibinfo {author} {\bibfnamefont {H.-J.}\ \bibnamefont {Werner}},\ }\bibfield  {title} {\bibinfo {title} {{MCSCF} optimization revisited. {II. C}ombined first- and second-order orbital optimization for large molecules},\ }\href {https://doi.org/10.1063/1.5142241} {\bibfield  {journal} {\bibinfo  {journal} {J. Chem. Phys.}\ }\textbf {\bibinfo {volume} {152}},\ \bibinfo {pages} {074102} (\bibinfo {year} {2020})}\BibitemShut {NoStop}%
\bibitem [{\citenamefont {Berning}\ \emph {et~al.}(2000)\citenamefont {Berning}, \citenamefont {Schweizer}, \citenamefont {Werner}, \citenamefont {Knowles},\ and\ \citenamefont {Palmieri}}]{so}%
  \BibitemOpen
  \bibfield  {author} {\bibinfo {author} {\bibfnamefont {A.}~\bibnamefont {Berning}}, \bibinfo {author} {\bibfnamefont {M.}~\bibnamefont {Schweizer}}, \bibinfo {author} {\bibfnamefont {H.-J.}\ \bibnamefont {Werner}}, \bibinfo {author} {\bibfnamefont {P.~J.}\ \bibnamefont {Knowles}},\ and\ \bibinfo {author} {\bibfnamefont {P.}~\bibnamefont {Palmieri}},\ }\bibfield  {title} {\bibinfo {title} {Spin-orbit matrix elements for internally contracted multireference configuration interaction wavefunctions},\ }\href {https://doi.org/10.1080/00268970009483386} {\bibfield  {journal} {\bibinfo  {journal} {Mol. Phys.}\ }\textbf {\bibinfo {volume} {98}},\ \bibinfo {pages} {1823} (\bibinfo {year} {2000})}\BibitemShut {NoStop}%
\bibitem [{\citenamefont {van~der Laan}\ and\ \citenamefont {Thole}(1988)}]{Laan1988}%
  \BibitemOpen
  \bibfield  {author} {\bibinfo {author} {\bibfnamefont {G.}~\bibnamefont {van~der Laan}}\ and\ \bibinfo {author} {\bibfnamefont {B.~T.}\ \bibnamefont {Thole}},\ }\bibfield  {title} {\bibinfo {title} {Local probe for spin-orbit interaction},\ }\href {https://doi.org/10.1103/physrevlett.60.1977} {\bibfield  {journal} {\bibinfo  {journal} {Physical Review Letters}\ }\textbf {\bibinfo {volume} {60}},\ \bibinfo {pages} {1977} (\bibinfo {year} {1988})}\BibitemShut {NoStop}%
\bibitem [{\citenamefont {Thole}\ and\ \citenamefont {van~der Laan}(1988{\natexlab{a}})}]{Thole1988}%
  \BibitemOpen
  \bibfield  {author} {\bibinfo {author} {\bibfnamefont {B.~T.}\ \bibnamefont {Thole}}\ and\ \bibinfo {author} {\bibfnamefont {G.}~\bibnamefont {van~der Laan}},\ }\bibfield  {title} {\bibinfo {title} {Branching ratio in x-ray absorption spectroscopy},\ }\href {https://doi.org/10.1103/physrevb.38.3158} {\bibfield  {journal} {\bibinfo  {journal} {Physical Review B}\ }\textbf {\bibinfo {volume} {38}},\ \bibinfo {pages} {3158} (\bibinfo {year} {1988}{\natexlab{a}})}\BibitemShut {NoStop}%
\bibitem [{\citenamefont {Thole}\ and\ \citenamefont {van~der Laan}(1988{\natexlab{b}})}]{Thole1988a}%
  \BibitemOpen
  \bibfield  {author} {\bibinfo {author} {\bibfnamefont {B.~T.}\ \bibnamefont {Thole}}\ and\ \bibinfo {author} {\bibfnamefont {G.}~\bibnamefont {van~der Laan}},\ }\bibfield  {title} {\bibinfo {title} {Linear relation between x-ray absorption branching ratio and valence-band spin-orbit expectation value},\ }\href {https://doi.org/10.1103/physreva.38.1943} {\bibfield  {journal} {\bibinfo  {journal} {Physical Review A}\ }\textbf {\bibinfo {volume} {38}},\ \bibinfo {pages} {1943} (\bibinfo {year} {1988}{\natexlab{b}})}\BibitemShut {NoStop}%
\bibitem [{\citenamefont {van~der Laan}\ \emph {et~al.}(2004)\citenamefont {van~der Laan}, \citenamefont {Moore}, \citenamefont {Tobin}, \citenamefont {Chung}, \citenamefont {Wall},\ and\ \citenamefont {Schwartz}}]{Laan2004}%
  \BibitemOpen
  \bibfield  {author} {\bibinfo {author} {\bibfnamefont {G.}~\bibnamefont {van~der Laan}}, \bibinfo {author} {\bibfnamefont {K.~T.}\ \bibnamefont {Moore}}, \bibinfo {author} {\bibfnamefont {J.~G.}\ \bibnamefont {Tobin}}, \bibinfo {author} {\bibfnamefont {B.~W.}\ \bibnamefont {Chung}}, \bibinfo {author} {\bibfnamefont {M.~A.}\ \bibnamefont {Wall}},\ and\ \bibinfo {author} {\bibfnamefont {A.~J.}\ \bibnamefont {Schwartz}},\ }\bibfield  {title} {\bibinfo {title} {Applicability of the spin-orbit sum rule for the actinide 5f states},\ }\href {https://doi.org/10.1103/physrevlett.93.097401} {\bibfield  {journal} {\bibinfo  {journal} {Physical Review Letters}\ }\textbf {\bibinfo {volume} {93}},\ \bibinfo {pages} {097401} (\bibinfo {year} {2004})}\BibitemShut {NoStop}%
\bibitem [{\citenamefont {Laguna-Marco}\ \emph {et~al.}(2010)\citenamefont {Laguna-Marco}, \citenamefont {Haskel}, \citenamefont {Souza-Neto}, \citenamefont {Lang}, \citenamefont {Krishnamurthy}, \citenamefont {Chikara}, \citenamefont {Cao},\ and\ \citenamefont {van Veenendaal}}]{LagunaMarco2010}%
  \BibitemOpen
  \bibfield  {author} {\bibinfo {author} {\bibfnamefont {M.~A.}\ \bibnamefont {Laguna-Marco}}, \bibinfo {author} {\bibfnamefont {D.}~\bibnamefont {Haskel}}, \bibinfo {author} {\bibfnamefont {N.}~\bibnamefont {Souza-Neto}}, \bibinfo {author} {\bibfnamefont {J.~C.}\ \bibnamefont {Lang}}, \bibinfo {author} {\bibfnamefont {V.~V.}\ \bibnamefont {Krishnamurthy}}, \bibinfo {author} {\bibfnamefont {S.}~\bibnamefont {Chikara}}, \bibinfo {author} {\bibfnamefont {G.}~\bibnamefont {Cao}},\ and\ \bibinfo {author} {\bibfnamefont {M.}~\bibnamefont {van Veenendaal}},\ }\bibfield  {title} {\bibinfo {title} {Orbital magnetism and spin-orbit effects in the electronic structure of {BaIrO}$_3$},\ }\href {https://doi.org/10.1103/physrevlett.105.216407} {\bibfield  {journal} {\bibinfo  {journal} {Physical Review Letters}\ }\textbf {\bibinfo {volume} {105}},\ \bibinfo {pages} {216407} (\bibinfo {year} {2010})}\BibitemShut {NoStop}%
\bibitem [{\citenamefont {Clancy}\ \emph {et~al.}(2012)\citenamefont {Clancy}, \citenamefont {Chen}, \citenamefont {Kim}, \citenamefont {Chen}, \citenamefont {Plumb}, \citenamefont {Jeon}, \citenamefont {Noh},\ and\ \citenamefont {Kim}}]{Clancy2012}%
  \BibitemOpen
  \bibfield  {author} {\bibinfo {author} {\bibfnamefont {J.~P.}\ \bibnamefont {Clancy}}, \bibinfo {author} {\bibfnamefont {N.}~\bibnamefont {Chen}}, \bibinfo {author} {\bibfnamefont {C.~Y.}\ \bibnamefont {Kim}}, \bibinfo {author} {\bibfnamefont {W.~F.}\ \bibnamefont {Chen}}, \bibinfo {author} {\bibfnamefont {K.~W.}\ \bibnamefont {Plumb}}, \bibinfo {author} {\bibfnamefont {B.~C.}\ \bibnamefont {Jeon}}, \bibinfo {author} {\bibfnamefont {T.~W.}\ \bibnamefont {Noh}},\ and\ \bibinfo {author} {\bibfnamefont {Y.-J.}\ \bibnamefont {Kim}},\ }\bibfield  {title} {\bibinfo {title} {Spin-orbit coupling in iridium-based $5d$ compounds probed by x-ray absorption spectroscopy},\ }\href {https://doi.org/10.1103/physrevb.86.195131} {\bibfield  {journal} {\bibinfo  {journal} {Physical Review B}\ }\textbf {\bibinfo {volume} {86}},\ \bibinfo {pages} {195131} (\bibinfo {year} {2012})}\BibitemShut {NoStop}%
\bibitem [{\citenamefont {Zwartsenberg}\ \emph {et~al.}(2020)\citenamefont {Zwartsenberg}, \citenamefont {Day}, \citenamefont {Razzoli}, \citenamefont {Michiardi}, \citenamefont {Xu}, \citenamefont {Shi}, \citenamefont {Denlinger}, \citenamefont {Cao}, \citenamefont {Calder}, \citenamefont {Ueda}, \citenamefont {Bertinshaw}, \citenamefont {Takagi}, \citenamefont {Kim}, \citenamefont {Elfimov},\ and\ \citenamefont {Damascelli}}]{Zwartsenberg2020}%
  \BibitemOpen
  \bibfield  {author} {\bibinfo {author} {\bibfnamefont {B.}~\bibnamefont {Zwartsenberg}}, \bibinfo {author} {\bibfnamefont {R.~P.}\ \bibnamefont {Day}}, \bibinfo {author} {\bibfnamefont {E.}~\bibnamefont {Razzoli}}, \bibinfo {author} {\bibfnamefont {M.}~\bibnamefont {Michiardi}}, \bibinfo {author} {\bibfnamefont {N.}~\bibnamefont {Xu}}, \bibinfo {author} {\bibfnamefont {M.}~\bibnamefont {Shi}}, \bibinfo {author} {\bibfnamefont {J.~D.}\ \bibnamefont {Denlinger}}, \bibinfo {author} {\bibfnamefont {G.}~\bibnamefont {Cao}}, \bibinfo {author} {\bibfnamefont {S.}~\bibnamefont {Calder}}, \bibinfo {author} {\bibfnamefont {K.}~\bibnamefont {Ueda}}, \bibinfo {author} {\bibfnamefont {J.}~\bibnamefont {Bertinshaw}}, \bibinfo {author} {\bibfnamefont {H.}~\bibnamefont {Takagi}}, \bibinfo {author} {\bibfnamefont {B.~J.}\ \bibnamefont {Kim}}, \bibinfo {author} {\bibfnamefont {I.~S.}\ \bibnamefont {Elfimov}},\ and\ \bibinfo {author} {\bibfnamefont {A.}~\bibnamefont {Damascelli}},\ }\bibfield  {title} {\bibinfo {title}
  {Spin-orbit-controlled metal–insulator transition in {Sr}$_2${IrO}$_4$},\ }\href {https://doi.org/10.1038/s41567-019-0750-y} {\bibfield  {journal} {\bibinfo  {journal} {Nature Physics}\ }\textbf {\bibinfo {volume} {16}},\ \bibinfo {pages} {290} (\bibinfo {year} {2020})}\BibitemShut {NoStop}%
\bibitem [{\citenamefont {Suzuki}\ \emph {et~al.}(2021)\citenamefont {Suzuki}, \citenamefont {Liu}, \citenamefont {Bertinshaw}, \citenamefont {Ueda}, \citenamefont {Kim}, \citenamefont {Laha}, \citenamefont {Weber}, \citenamefont {Yang}, \citenamefont {Wang}, \citenamefont {Takahashi}, \citenamefont {Fürsich}, \citenamefont {Minola}, \citenamefont {Lotsch}, \citenamefont {Kim}, \citenamefont {Yavaş}, \citenamefont {Daghofer}, \citenamefont {Chaloupka}, \citenamefont {Khaliullin}, \citenamefont {Gretarsson},\ and\ \citenamefont {Keimer}}]{Suzuki2021}%
  \BibitemOpen
  \bibfield  {author} {\bibinfo {author} {\bibfnamefont {H.}~\bibnamefont {Suzuki}}, \bibinfo {author} {\bibfnamefont {H.}~\bibnamefont {Liu}}, \bibinfo {author} {\bibfnamefont {J.}~\bibnamefont {Bertinshaw}}, \bibinfo {author} {\bibfnamefont {K.}~\bibnamefont {Ueda}}, \bibinfo {author} {\bibfnamefont {H.}~\bibnamefont {Kim}}, \bibinfo {author} {\bibfnamefont {S.}~\bibnamefont {Laha}}, \bibinfo {author} {\bibfnamefont {D.}~\bibnamefont {Weber}}, \bibinfo {author} {\bibfnamefont {Z.}~\bibnamefont {Yang}}, \bibinfo {author} {\bibfnamefont {L.}~\bibnamefont {Wang}}, \bibinfo {author} {\bibfnamefont {H.}~\bibnamefont {Takahashi}}, \bibinfo {author} {\bibfnamefont {K.}~\bibnamefont {Fürsich}}, \bibinfo {author} {\bibfnamefont {M.}~\bibnamefont {Minola}}, \bibinfo {author} {\bibfnamefont {B.~V.}\ \bibnamefont {Lotsch}}, \bibinfo {author} {\bibfnamefont {B.~J.}\ \bibnamefont {Kim}}, \bibinfo {author} {\bibfnamefont {H.}~\bibnamefont {Yavaş}}, \bibinfo {author} {\bibfnamefont {M.}~\bibnamefont {Daghofer}}, \bibinfo
  {author} {\bibfnamefont {J.}~\bibnamefont {Chaloupka}}, \bibinfo {author} {\bibfnamefont {G.}~\bibnamefont {Khaliullin}}, \bibinfo {author} {\bibfnamefont {H.}~\bibnamefont {Gretarsson}},\ and\ \bibinfo {author} {\bibfnamefont {B.}~\bibnamefont {Keimer}},\ }\bibfield  {title} {\bibinfo {title} {Proximate ferromagnetic state in the kitaev model material $\alpha$-{RuCl}$_3$},\ }\href {https://doi.org/10.1038/s41467-021-24722-4} {\bibfield  {journal} {\bibinfo  {journal} {Nature Communications}\ }\textbf {\bibinfo {volume} {12}},\ \bibinfo {pages} {4512} (\bibinfo {year} {2021})}\BibitemShut {NoStop}%
\bibitem [{\citenamefont {Lamichhane}\ \emph {et~al.}(2024)\citenamefont {Lamichhane}, \citenamefont {Sankhi}, \citenamefont {Kundu}, \citenamefont {Fabbris}, \citenamefont {Choi}, \citenamefont {Haskel}, \citenamefont {McChesney}, \citenamefont {Cao}, \citenamefont {Li}, \citenamefont {Bisogni}, \citenamefont {Borunda},\ and\ \citenamefont {Meyers}}]{Lamichhane2024}%
  \BibitemOpen
  \bibfield  {author} {\bibinfo {author} {\bibfnamefont {U.}~\bibnamefont {Lamichhane}}, \bibinfo {author} {\bibfnamefont {B.}~\bibnamefont {Sankhi}}, \bibinfo {author} {\bibfnamefont {N.}~\bibnamefont {Kundu}}, \bibinfo {author} {\bibfnamefont {G.}~\bibnamefont {Fabbris}}, \bibinfo {author} {\bibfnamefont {Y.}~\bibnamefont {Choi}}, \bibinfo {author} {\bibfnamefont {D.}~\bibnamefont {Haskel}}, \bibinfo {author} {\bibfnamefont {J.~L.}\ \bibnamefont {McChesney}}, \bibinfo {author} {\bibfnamefont {Y.}~\bibnamefont {Cao}}, \bibinfo {author} {\bibfnamefont {J.}~\bibnamefont {Li}}, \bibinfo {author} {\bibfnamefont {V.}~\bibnamefont {Bisogni}}, \bibinfo {author} {\bibfnamefont {M.~F.}\ \bibnamefont {Borunda}},\ and\ \bibinfo {author} {\bibfnamefont {D.}~\bibnamefont {Meyers}},\ }\bibfield  {title} {\bibinfo {title} {Electronic reconstruction in confined {SrRuO}$_3$ monolayers},\ }\href {https://doi.org/10.1103/physrevb.110.235104} {\bibfield  {journal} {\bibinfo  {journal} {Physical Review B}\ }\textbf {\bibinfo
  {volume} {110}},\ \bibinfo {pages} {235104} (\bibinfo {year} {2024})}\BibitemShut {NoStop}%
\bibitem [{\citenamefont {Bahrami}\ \emph {et~al.}(2022)\citenamefont {Bahrami}, \citenamefont {Hu}, \citenamefont {Du}, \citenamefont {Lebedev}, \citenamefont {Wang}, \citenamefont {Luetkens}, \citenamefont {Fabbris}, \citenamefont {Graf}, \citenamefont {Haskel}, \citenamefont {Ran},\ and\ \citenamefont {Tafti}}]{Bahrami2022}%
  \BibitemOpen
  \bibfield  {author} {\bibinfo {author} {\bibfnamefont {F.}~\bibnamefont {Bahrami}}, \bibinfo {author} {\bibfnamefont {X.}~\bibnamefont {Hu}}, \bibinfo {author} {\bibfnamefont {Y.}~\bibnamefont {Du}}, \bibinfo {author} {\bibfnamefont {O.~I.}\ \bibnamefont {Lebedev}}, \bibinfo {author} {\bibfnamefont {C.}~\bibnamefont {Wang}}, \bibinfo {author} {\bibfnamefont {H.}~\bibnamefont {Luetkens}}, \bibinfo {author} {\bibfnamefont {G.}~\bibnamefont {Fabbris}}, \bibinfo {author} {\bibfnamefont {M.~J.}\ \bibnamefont {Graf}}, \bibinfo {author} {\bibfnamefont {D.}~\bibnamefont {Haskel}}, \bibinfo {author} {\bibfnamefont {Y.}~\bibnamefont {Ran}},\ and\ \bibinfo {author} {\bibfnamefont {F.}~\bibnamefont {Tafti}},\ }\bibfield  {title} {\bibinfo {title} {First demonstration of tuning between the {Kitaev} and {Ising} limits in a honeycomb lattice},\ }\href {https://doi.org/10.1126/sciadv.abl5671} {\bibfield  {journal} {\bibinfo  {journal} {Science Advances}\ }\textbf {\bibinfo {volume} {8}},\ \bibinfo {pages} {eabl5671}
  (\bibinfo {year} {2022})}\BibitemShut {NoStop}%
\bibitem [{\citenamefont {Lebert}\ \emph {et~al.}(2020)\citenamefont {Lebert}, \citenamefont {Kim}, \citenamefont {Bisogni}, \citenamefont {Jarrige}, \citenamefont {Barbour},\ and\ \citenamefont {Kim}}]{M_edge}%
  \BibitemOpen
  \bibfield  {author} {\bibinfo {author} {\bibfnamefont {B.~W.}\ \bibnamefont {Lebert}}, \bibinfo {author} {\bibfnamefont {S.}~\bibnamefont {Kim}}, \bibinfo {author} {\bibfnamefont {V.}~\bibnamefont {Bisogni}}, \bibinfo {author} {\bibfnamefont {I.}~\bibnamefont {Jarrige}}, \bibinfo {author} {\bibfnamefont {A.~M.}\ \bibnamefont {Barbour}},\ and\ \bibinfo {author} {\bibfnamefont {Y.-J.}\ \bibnamefont {Kim}},\ }\bibfield  {title} {\bibinfo {title} {Resonant inelastic x-ray scattering study of $\alpha$-{RuCl}$_3$: a progress report},\ }\href {https://doi.org/10.1088/1361-648X/ab5595} {\bibfield  {journal} {\bibinfo  {journal} {J. Phys.: Condens. Matter}\ }\textbf {\bibinfo {volume} {32}},\ \bibinfo {pages} {144001} (\bibinfo {year} {2020})}\BibitemShut {NoStop}%
\bibitem [{\citenamefont {Zhang}\ \emph {et~al.}(2022)\citenamefont {Zhang}, \citenamefont {Cheng}, \citenamefont {Cao}, \citenamefont {Bao}, \citenamefont {Zhao}, \citenamefont {Liu}, \citenamefont {Zhao}, \citenamefont {Choi}, \citenamefont {Zhou}, \citenamefont {Shafer},\ and\ \citenamefont {Zhai}}]{Zhang2022}%
  \BibitemOpen
  \bibfield  {author} {\bibinfo {author} {\bibfnamefont {J.}~\bibnamefont {Zhang}}, \bibinfo {author} {\bibfnamefont {L.}~\bibnamefont {Cheng}}, \bibinfo {author} {\bibfnamefont {H.}~\bibnamefont {Cao}}, \bibinfo {author} {\bibfnamefont {M.}~\bibnamefont {Bao}}, \bibinfo {author} {\bibfnamefont {J.}~\bibnamefont {Zhao}}, \bibinfo {author} {\bibfnamefont {X.}~\bibnamefont {Liu}}, \bibinfo {author} {\bibfnamefont {A.}~\bibnamefont {Zhao}}, \bibinfo {author} {\bibfnamefont {Y.}~\bibnamefont {Choi}}, \bibinfo {author} {\bibfnamefont {H.}~\bibnamefont {Zhou}}, \bibinfo {author} {\bibfnamefont {P.}~\bibnamefont {Shafer}},\ and\ \bibinfo {author} {\bibfnamefont {X.}~\bibnamefont {Zhai}},\ }\bibfield  {title} {\bibinfo {title} {The exceedingly strong two-dimensional ferromagnetism in bi-atomic layer {SrRuO}$_3$ with a critical conduction transition},\ }\href {https://doi.org/10.1007/s12274-022-4392-5} {\bibfield  {journal} {\bibinfo  {journal} {Nano Research}\ }\textbf {\bibinfo {volume} {15}},\ \bibinfo {pages}
  {7584} (\bibinfo {year} {2022})}\BibitemShut {NoStop}%
\bibitem [{\citenamefont {Takahashi}\ \emph {et~al.}(2021)\citenamefont {Takahashi}, \citenamefont {Suzuki}, \citenamefont {Bertinshaw}, \citenamefont {Bette}, \citenamefont {Muhle}, \citenamefont {Nuss}, \citenamefont {Dinnebier}, \citenamefont {Yaresko}, \citenamefont {Khaliullin}, \citenamefont {Gretarsson}, \citenamefont {Takayama}, \citenamefont {Takagi},\ and\ \citenamefont {Keimer}}]{Takahashi2021}%
  \BibitemOpen
  \bibfield  {author} {\bibinfo {author} {\bibfnamefont {H.}~\bibnamefont {Takahashi}}, \bibinfo {author} {\bibfnamefont {H.}~\bibnamefont {Suzuki}}, \bibinfo {author} {\bibfnamefont {J.}~\bibnamefont {Bertinshaw}}, \bibinfo {author} {\bibfnamefont {S.}~\bibnamefont {Bette}}, \bibinfo {author} {\bibfnamefont {C.}~\bibnamefont {Muhle}}, \bibinfo {author} {\bibfnamefont {J.}~\bibnamefont {Nuss}}, \bibinfo {author} {\bibfnamefont {R.}~\bibnamefont {Dinnebier}}, \bibinfo {author} {\bibfnamefont {A.}~\bibnamefont {Yaresko}}, \bibinfo {author} {\bibfnamefont {G.}~\bibnamefont {Khaliullin}}, \bibinfo {author} {\bibfnamefont {H.}~\bibnamefont {Gretarsson}}, \bibinfo {author} {\bibfnamefont {T.}~\bibnamefont {Takayama}}, \bibinfo {author} {\bibfnamefont {H.}~\bibnamefont {Takagi}},\ and\ \bibinfo {author} {\bibfnamefont {B.}~\bibnamefont {Keimer}},\ }\bibfield  {title} {\bibinfo {title} {Nonmagnetic ${J}=0$ state and spin-orbit excitations in {K}$_2${RuCl}$_6$},\ }\href {https://doi.org/10.1103/physrevlett.127.227201}
  {\bibfield  {journal} {\bibinfo  {journal} {Physical Review Letters}\ }\textbf {\bibinfo {volume} {127}},\ \bibinfo {pages} {227201} (\bibinfo {year} {2021})}\BibitemShut {NoStop}%
\bibitem [{\citenamefont {Gretarsson}\ \emph {et~al.}(2024)\citenamefont {Gretarsson}, \citenamefont {Fujihara}, \citenamefont {Sato}, \citenamefont {Gotou}, \citenamefont {Imai}, \citenamefont {Ohgushi}, \citenamefont {Keimer},\ and\ \citenamefont {Suzuki}}]{Gretarsson2024}%
  \BibitemOpen
  \bibfield  {author} {\bibinfo {author} {\bibfnamefont {H.}~\bibnamefont {Gretarsson}}, \bibinfo {author} {\bibfnamefont {H.}~\bibnamefont {Fujihara}}, \bibinfo {author} {\bibfnamefont {F.}~\bibnamefont {Sato}}, \bibinfo {author} {\bibfnamefont {H.}~\bibnamefont {Gotou}}, \bibinfo {author} {\bibfnamefont {Y.}~\bibnamefont {Imai}}, \bibinfo {author} {\bibfnamefont {K.}~\bibnamefont {Ohgushi}}, \bibinfo {author} {\bibfnamefont {B.}~\bibnamefont {Keimer}},\ and\ \bibinfo {author} {\bibfnamefont {H.}~\bibnamefont {Suzuki}},\ }\bibfield  {title} {\bibinfo {title} {${J} = 1/2$ pseudospins and $d$-$p$ hybridization in the {Kitaev} spin liquid candidates {RuX}$_3$ ({X} = {Cl, Br, I})},\ }\href {https://doi.org/10.1103/physrevb.109.l180413} {\bibfield  {journal} {\bibinfo  {journal} {Physical Review B}\ }\textbf {\bibinfo {volume} {109}},\ \bibinfo {pages} {l180413} (\bibinfo {year} {2024})}\BibitemShut {NoStop}%
\bibitem [{\citenamefont {Abdeldaim}\ \emph {et~al.}(2024)\citenamefont {Abdeldaim}, \citenamefont {Gretarsson}, \citenamefont {Day}, \citenamefont {Le}, \citenamefont {Stenning}, \citenamefont {Manuel}, \citenamefont {Perry}, \citenamefont {Tsirlin}, \citenamefont {Nilsen},\ and\ \citenamefont {Clark}}]{Abdeldaim2024}%
  \BibitemOpen
  \bibfield  {author} {\bibinfo {author} {\bibfnamefont {A.~H.}\ \bibnamefont {Abdeldaim}}, \bibinfo {author} {\bibfnamefont {H.}~\bibnamefont {Gretarsson}}, \bibinfo {author} {\bibfnamefont {S.~J.}\ \bibnamefont {Day}}, \bibinfo {author} {\bibfnamefont {M.~D.}\ \bibnamefont {Le}}, \bibinfo {author} {\bibfnamefont {G.~B.~G.}\ \bibnamefont {Stenning}}, \bibinfo {author} {\bibfnamefont {P.}~\bibnamefont {Manuel}}, \bibinfo {author} {\bibfnamefont {R.~S.}\ \bibnamefont {Perry}}, \bibinfo {author} {\bibfnamefont {A.~A.}\ \bibnamefont {Tsirlin}}, \bibinfo {author} {\bibfnamefont {G.~J.}\ \bibnamefont {Nilsen}},\ and\ \bibinfo {author} {\bibfnamefont {L.}~\bibnamefont {Clark}},\ }\bibfield  {title} {\bibinfo {title} {Kitaev interactions through extended superexchange pathways in the ${J}_{eff}=1/2$ {Ru}$^{3+}$ honeycomb magnet {RuP}$_3${SiO}$_{11}$},\ }\href {https://doi.org/10.1038/s41467-024-53900-3} {\bibfield  {journal} {\bibinfo  {journal} {Nature Communications}\ }\textbf {\bibinfo {volume} {15}},\ \bibinfo
  {pages} {9778} (\bibinfo {year} {2024})}\BibitemShut {NoStop}%
\bibitem [{\citenamefont {Bertinshaw}\ \emph {et~al.}(2021)\citenamefont {Bertinshaw}, \citenamefont {Krautloher}, \citenamefont {Suzuki}, \citenamefont {Takahashi}, \citenamefont {Ivanov}, \citenamefont {Yavas}, \citenamefont {Kim}, \citenamefont {Gretarsson},\ and\ \citenamefont {Keimer}}]{Bertinshaw2021}%
  \BibitemOpen
  \bibfield  {author} {\bibinfo {author} {\bibfnamefont {J.}~\bibnamefont {Bertinshaw}}, \bibinfo {author} {\bibfnamefont {M.}~\bibnamefont {Krautloher}}, \bibinfo {author} {\bibfnamefont {H.}~\bibnamefont {Suzuki}}, \bibinfo {author} {\bibfnamefont {H.}~\bibnamefont {Takahashi}}, \bibinfo {author} {\bibfnamefont {A.}~\bibnamefont {Ivanov}}, \bibinfo {author} {\bibfnamefont {H.}~\bibnamefont {Yavas}}, \bibinfo {author} {\bibfnamefont {B.~J.}\ \bibnamefont {Kim}}, \bibinfo {author} {\bibfnamefont {H.}~\bibnamefont {Gretarsson}},\ and\ \bibinfo {author} {\bibfnamefont {B.}~\bibnamefont {Keimer}},\ }\bibfield  {title} {\bibinfo {title} {Spin and charge excitations in the correlated multiband metal {Ca}$_3${Ru}$_2${O}$_7$},\ }\href {https://doi.org/10.1103/physrevb.103.085108} {\bibfield  {journal} {\bibinfo  {journal} {Physical Review B}\ }\textbf {\bibinfo {volume} {103}},\ \bibinfo {pages} {085108} (\bibinfo {year} {2021})}\BibitemShut {NoStop}%
\bibitem [{\citenamefont {Suzuki}\ \emph {et~al.}(2023)\citenamefont {Suzuki}, \citenamefont {Wang}, \citenamefont {Bertinshaw}, \citenamefont {Strand}, \citenamefont {Kaser}, \citenamefont {Krautloher}, \citenamefont {Yang}, \citenamefont {Wentzell}, \citenamefont {Parcollet}, \citenamefont {Jerzembeck}, \citenamefont {Kikugawa}, \citenamefont {Mackenzie}, \citenamefont {Georges}, \citenamefont {Hansmann}, \citenamefont {Gretarsson},\ and\ \citenamefont {Keimer}}]{Suzuki2023}%
  \BibitemOpen
  \bibfield  {author} {\bibinfo {author} {\bibfnamefont {H.}~\bibnamefont {Suzuki}}, \bibinfo {author} {\bibfnamefont {L.}~\bibnamefont {Wang}}, \bibinfo {author} {\bibfnamefont {J.}~\bibnamefont {Bertinshaw}}, \bibinfo {author} {\bibfnamefont {H.~U.~R.}\ \bibnamefont {Strand}}, \bibinfo {author} {\bibfnamefont {S.}~\bibnamefont {Kaser}}, \bibinfo {author} {\bibfnamefont {M.}~\bibnamefont {Krautloher}}, \bibinfo {author} {\bibfnamefont {Z.}~\bibnamefont {Yang}}, \bibinfo {author} {\bibfnamefont {N.}~\bibnamefont {Wentzell}}, \bibinfo {author} {\bibfnamefont {O.}~\bibnamefont {Parcollet}}, \bibinfo {author} {\bibfnamefont {F.}~\bibnamefont {Jerzembeck}}, \bibinfo {author} {\bibfnamefont {N.}~\bibnamefont {Kikugawa}}, \bibinfo {author} {\bibfnamefont {A.~P.}\ \bibnamefont {Mackenzie}}, \bibinfo {author} {\bibfnamefont {A.}~\bibnamefont {Georges}}, \bibinfo {author} {\bibfnamefont {P.}~\bibnamefont {Hansmann}}, \bibinfo {author} {\bibfnamefont {H.}~\bibnamefont {Gretarsson}},\ and\ \bibinfo {author}
  {\bibfnamefont {B.}~\bibnamefont {Keimer}},\ }\bibfield  {title} {\bibinfo {title} {Distinct spin and orbital dynamics in {Sr}$_2${RuO}$_4$},\ }\href {https://doi.org/10.1038/s41467-023-42804-3} {\bibfield  {journal} {\bibinfo  {journal} {Nature Communications}\ }\textbf {\bibinfo {volume} {14}},\ \bibinfo {pages} {7042} (\bibinfo {year} {2023})}\BibitemShut {NoStop}%
\bibitem [{\citenamefont {Gretarsson}\ \emph {et~al.}(2019)\citenamefont {Gretarsson}, \citenamefont {Suzuki}, \citenamefont {Kim}, \citenamefont {Ueda}, \citenamefont {Krautloher}, \citenamefont {Kim}, \citenamefont {Yavas}, \citenamefont {Khaliullin},\ and\ \citenamefont {Keimer}}]{Gretarsson2019}%
  \BibitemOpen
  \bibfield  {author} {\bibinfo {author} {\bibfnamefont {H.}~\bibnamefont {Gretarsson}}, \bibinfo {author} {\bibfnamefont {H.}~\bibnamefont {Suzuki}}, \bibinfo {author} {\bibfnamefont {H.}~\bibnamefont {Kim}}, \bibinfo {author} {\bibfnamefont {K.}~\bibnamefont {Ueda}}, \bibinfo {author} {\bibfnamefont {M.}~\bibnamefont {Krautloher}}, \bibinfo {author} {\bibfnamefont {B.~J.}\ \bibnamefont {Kim}}, \bibinfo {author} {\bibfnamefont {H.}~\bibnamefont {Yavas}}, \bibinfo {author} {\bibfnamefont {G.}~\bibnamefont {Khaliullin}},\ and\ \bibinfo {author} {\bibfnamefont {B.}~\bibnamefont {Keimer}},\ }\bibfield  {title} {\bibinfo {title} {Observation of spin-orbit excitations and {Hund's} multiplets in {Ca}$_2${RuO}$_4$},\ }\href {https://doi.org/10.1103/physrevb.100.045123} {\bibfield  {journal} {\bibinfo  {journal} {Physical Review B}\ }\textbf {\bibinfo {volume} {100}},\ \bibinfo {pages} {045123} (\bibinfo {year} {2019})}\BibitemShut {NoStop}%
\bibitem [{\citenamefont {Kim}\ \emph {et~al.}(2012)\citenamefont {Kim}, \citenamefont {Casa}, \citenamefont {Upton}, \citenamefont {Gog}, \citenamefont {Kim}, \citenamefont {Mitchell}, \citenamefont {van Veenendaal}, \citenamefont {Daghofer}, \citenamefont {van~den Brink}, \citenamefont {Khaliullin},\ and\ \citenamefont {Kim}}]{kim2012}%
  \BibitemOpen
  \bibfield  {author} {\bibinfo {author} {\bibfnamefont {J.}~\bibnamefont {Kim}}, \bibinfo {author} {\bibfnamefont {D.}~\bibnamefont {Casa}}, \bibinfo {author} {\bibfnamefont {M.~H.}\ \bibnamefont {Upton}}, \bibinfo {author} {\bibfnamefont {T.}~\bibnamefont {Gog}}, \bibinfo {author} {\bibfnamefont {Y.-J.}\ \bibnamefont {Kim}}, \bibinfo {author} {\bibfnamefont {J.~F.}\ \bibnamefont {Mitchell}}, \bibinfo {author} {\bibfnamefont {M.}~\bibnamefont {van Veenendaal}}, \bibinfo {author} {\bibfnamefont {M.}~\bibnamefont {Daghofer}}, \bibinfo {author} {\bibfnamefont {J.}~\bibnamefont {van~den Brink}}, \bibinfo {author} {\bibfnamefont {G.}~\bibnamefont {Khaliullin}},\ and\ \bibinfo {author} {\bibfnamefont {B.~J.}\ \bibnamefont {Kim}},\ }\bibfield  {title} {\bibinfo {title} {Magnetic excitation spectra of {Sr}$_2${IrO}$_4$ probed by resonant inelastic x-ray scattering: Establishing links to cuprate superconductors},\ }\href {https://doi.org/10.1103/PhysRevLett.108.177003} {\bibfield  {journal} {\bibinfo  {journal}
  {Phys. Rev. Lett.}\ }\textbf {\bibinfo {volume} {108}},\ \bibinfo {pages} {177003} (\bibinfo {year} {2012})}\BibitemShut {NoStop}%
\bibitem [{\citenamefont {Imai}\ \emph {et~al.}(2022)\citenamefont {Imai}, \citenamefont {Nawa}, \citenamefont {Shimizu}, \citenamefont {Yamada}, \citenamefont {Fujihara}, \citenamefont {Aoyama}, \citenamefont {Takahashi}, \citenamefont {Okuyama}, \citenamefont {Ohashi}, \citenamefont {Hagihala}, \citenamefont {Torii}, \citenamefont {Morikawa}, \citenamefont {Terauchi}, \citenamefont {Kawamata}, \citenamefont {Kato}, \citenamefont {Gotou}, \citenamefont {Itoh}, \citenamefont {Sato},\ and\ \citenamefont {Ohgushi}}]{Imai2022}%
  \BibitemOpen
  \bibfield  {author} {\bibinfo {author} {\bibfnamefont {Y.}~\bibnamefont {Imai}}, \bibinfo {author} {\bibfnamefont {K.}~\bibnamefont {Nawa}}, \bibinfo {author} {\bibfnamefont {Y.}~\bibnamefont {Shimizu}}, \bibinfo {author} {\bibfnamefont {W.}~\bibnamefont {Yamada}}, \bibinfo {author} {\bibfnamefont {H.}~\bibnamefont {Fujihara}}, \bibinfo {author} {\bibfnamefont {T.}~\bibnamefont {Aoyama}}, \bibinfo {author} {\bibfnamefont {R.}~\bibnamefont {Takahashi}}, \bibinfo {author} {\bibfnamefont {D.}~\bibnamefont {Okuyama}}, \bibinfo {author} {\bibfnamefont {T.}~\bibnamefont {Ohashi}}, \bibinfo {author} {\bibfnamefont {M.}~\bibnamefont {Hagihala}}, \bibinfo {author} {\bibfnamefont {S.}~\bibnamefont {Torii}}, \bibinfo {author} {\bibfnamefont {D.}~\bibnamefont {Morikawa}}, \bibinfo {author} {\bibfnamefont {M.}~\bibnamefont {Terauchi}}, \bibinfo {author} {\bibfnamefont {T.}~\bibnamefont {Kawamata}}, \bibinfo {author} {\bibfnamefont {M.}~\bibnamefont {Kato}}, \bibinfo {author} {\bibfnamefont {H.}~\bibnamefont {Gotou}},
  \bibinfo {author} {\bibfnamefont {M.}~\bibnamefont {Itoh}}, \bibinfo {author} {\bibfnamefont {T.~J.}\ \bibnamefont {Sato}},\ and\ \bibinfo {author} {\bibfnamefont {K.}~\bibnamefont {Ohgushi}},\ }\bibfield  {title} {\bibinfo {title} {Zigzag magnetic order in the kitaev spin-liquid candidate material {RuBr}$_3$ with a honeycomb lattice},\ }\href {https://doi.org/10.1103/physrevb.105.l041112} {\bibfield  {journal} {\bibinfo  {journal} {Physical Review B}\ }\textbf {\bibinfo {volume} {105}},\ \bibinfo {pages} {l041112} (\bibinfo {year} {2022})}\BibitemShut {NoStop}%
\bibitem [{\citenamefont {Nawa}\ \emph {et~al.}(2021)\citenamefont {Nawa}, \citenamefont {Imai}, \citenamefont {Yamaji}, \citenamefont {Fujihara}, \citenamefont {Yamada}, \citenamefont {Takahashi}, \citenamefont {Hiraoka}, \citenamefont {Hagihala}, \citenamefont {Torii}, \citenamefont {Aoyama}, \citenamefont {Ohashi}, \citenamefont {Shimizu}, \citenamefont {Gotou}, \citenamefont {Itoh}, \citenamefont {Ohgushi},\ and\ \citenamefont {Sato}}]{Nawa2021}%
  \BibitemOpen
  \bibfield  {author} {\bibinfo {author} {\bibfnamefont {K.}~\bibnamefont {Nawa}}, \bibinfo {author} {\bibfnamefont {Y.}~\bibnamefont {Imai}}, \bibinfo {author} {\bibfnamefont {Y.}~\bibnamefont {Yamaji}}, \bibinfo {author} {\bibfnamefont {H.}~\bibnamefont {Fujihara}}, \bibinfo {author} {\bibfnamefont {W.}~\bibnamefont {Yamada}}, \bibinfo {author} {\bibfnamefont {R.}~\bibnamefont {Takahashi}}, \bibinfo {author} {\bibfnamefont {T.}~\bibnamefont {Hiraoka}}, \bibinfo {author} {\bibfnamefont {M.}~\bibnamefont {Hagihala}}, \bibinfo {author} {\bibfnamefont {S.}~\bibnamefont {Torii}}, \bibinfo {author} {\bibfnamefont {T.}~\bibnamefont {Aoyama}}, \bibinfo {author} {\bibfnamefont {T.}~\bibnamefont {Ohashi}}, \bibinfo {author} {\bibfnamefont {Y.}~\bibnamefont {Shimizu}}, \bibinfo {author} {\bibfnamefont {H.}~\bibnamefont {Gotou}}, \bibinfo {author} {\bibfnamefont {M.}~\bibnamefont {Itoh}}, \bibinfo {author} {\bibfnamefont {K.}~\bibnamefont {Ohgushi}},\ and\ \bibinfo {author} {\bibfnamefont {T.~J.}\ \bibnamefont
  {Sato}},\ }\bibfield  {title} {\bibinfo {title} {Strongly electron-correlated semimetal {RuI}$_3$ with a layered honeycomb structure},\ }\href {https://doi.org/10.7566/jpsj.90.123703} {\bibfield  {journal} {\bibinfo  {journal} {Journal of the Physical Society of Japan}\ }\textbf {\bibinfo {volume} {90}},\ \bibinfo {pages} {123703} (\bibinfo {year} {2021})}\BibitemShut {NoStop}%
\bibitem [{\citenamefont {Ni}\ \emph {et~al.}(2022)\citenamefont {Ni}, \citenamefont {Gui}, \citenamefont {Powderly},\ and\ \citenamefont {Cava}}]{Ni2022}%
  \BibitemOpen
  \bibfield  {author} {\bibinfo {author} {\bibfnamefont {D.}~\bibnamefont {Ni}}, \bibinfo {author} {\bibfnamefont {X.}~\bibnamefont {Gui}}, \bibinfo {author} {\bibfnamefont {K.~M.}\ \bibnamefont {Powderly}},\ and\ \bibinfo {author} {\bibfnamefont {R.~J.}\ \bibnamefont {Cava}},\ }\bibfield  {title} {\bibinfo {title} {Honeycomb‐structure {RuI}$_3$, a new quantum material related to $\alpha$-{RuCl}$_3$},\ }\href {https://doi.org/10.1002/adma.202106831} {\bibfield  {journal} {\bibinfo  {journal} {Advanced Materials}\ }\textbf {\bibinfo {volume} {34}},\ \bibinfo {pages} {2106831} (\bibinfo {year} {2022})}\BibitemShut {NoStop}%
\bibitem [{\citenamefont {Sato}\ \emph {et~al.}(2024)\citenamefont {Sato}, \citenamefont {Fujihara}, \citenamefont {Gotou}, \citenamefont {Aoyama}, \citenamefont {Imai},\ and\ \citenamefont {Ohgushi}}]{Sato2024}%
  \BibitemOpen
  \bibfield  {author} {\bibinfo {author} {\bibfnamefont {F.}~\bibnamefont {Sato}}, \bibinfo {author} {\bibfnamefont {H.}~\bibnamefont {Fujihara}}, \bibinfo {author} {\bibfnamefont {H.}~\bibnamefont {Gotou}}, \bibinfo {author} {\bibfnamefont {T.}~\bibnamefont {Aoyama}}, \bibinfo {author} {\bibfnamefont {Y.}~\bibnamefont {Imai}},\ and\ \bibinfo {author} {\bibfnamefont {K.}~\bibnamefont {Ohgushi}},\ }\bibfield  {title} {\bibinfo {title} {Insulator-metal transition in {Ru}({Br}$_{1-x}${I}$_x$)$_3$ with honeycomb structure},\ }\href {https://doi.org/10.1103/physrevb.109.035154} {\bibfield  {journal} {\bibinfo  {journal} {Physical Review B}\ }\textbf {\bibinfo {volume} {109}},\ \bibinfo {pages} {035154} (\bibinfo {year} {2024})}\BibitemShut {NoStop}%
\bibitem [{\citenamefont {Kanoda}\ and\ \citenamefont {Kato}(2011)}]{Kanoda2011}%
  \BibitemOpen
  \bibfield  {author} {\bibinfo {author} {\bibfnamefont {K.}~\bibnamefont {Kanoda}}\ and\ \bibinfo {author} {\bibfnamefont {R.}~\bibnamefont {Kato}},\ }\bibfield  {title} {\bibinfo {title} {{Mott} physics in organic conductors with triangular lattices},\ }\href {https://doi.org/10.1146/annurev-conmatphys-062910-140521} {\bibfield  {journal} {\bibinfo  {journal} {Annual Review of Condensed Matter Physics}\ }\textbf {\bibinfo {volume} {2}},\ \bibinfo {pages} {167} (\bibinfo {year} {2011})}\BibitemShut {NoStop}%
\bibitem [{\citenamefont {Tam}(2026)}]{zenodo}%
  \BibitemOpen
  \bibfield  {author} {\bibinfo {author} {\bibfnamefont {C.~C.}\ \bibnamefont {Tam}},\ }\bibfield  {title} {\bibinfo {title} {Near itinerancy and slow singlet formation in the triangular lattice {NaRuO}$_2$ [{Dataset}]},\ }\bibfield  {journal} {\bibinfo  {journal} {Zenodo}\ }\href {https://doi.org/10.5281/zenodo.21147485} {10.5281/zenodo.21147485} (\bibinfo {year} {2026})\BibitemShut {NoStop}%
\end{thebibliography}
\end{document}